\documentclass[11pt]{article}
\pdfoutput=1
\usepackage[utf8]{inputenc}
\usepackage[english]{babel}
\usepackage[T1]{fontenc}
\usepackage{lmodern}
\usepackage[letterpaper,margin=1in]{geometry}
\usepackage{microtype}
\usepackage{amsmath,amssymb,amsthm}
\usepackage{braket}
\usepackage{mathtools}
\usepackage{float}
\usepackage{xcolor}
\usepackage{booktabs}
\usepackage{graphicx}
\usepackage{tikz}
\usepackage{tcolorbox}
\usepackage[numbers,sort&compress]{natbib}
\usepackage{hyperref}
\hypersetup{colorlinks=true,linkcolor=blue!60!black,citecolor=green!45!black,urlcolor=blue!60!black}

\tcbuselibrary{skins,breakable}

\theoremstyle{plain}
\newtheorem{theorem}{Theorem}[section]
\newtheorem{lemma}[theorem]{Lemma}
\newtheorem{proposition}[theorem]{Proposition}
\newtheorem{corollary}[theorem]{Corollary}
\newtheorem{assumption}[theorem]{Assumption}
\theoremstyle{definition}
\newtheorem{definition}[theorem]{Definition}
\theoremstyle{remark}
\newtheorem{remark}[theorem]{Remark}

\newtcolorbox{protocolbox}[1][]{%
  colback=gray!8, colframe=gray!60, coltitle=black,
  fonttitle=\bfseries, boxrule=0.8pt, arc=2pt,
  left=8pt, right=8pt, top=8pt, bottom=8pt,
  before skip=12pt, after skip=12pt, breakable, #1
}
\newcounter{protocol}

\newcommand{\Tr}{\mathrm{Tr}}
\newcommand{\poly}{\mathrm{poly}}
\newcommand{\negl}{\mathrm{negl}}

\newcommand{\id}{\mathbf{I}}

\newcommand{\E}{\mathbb{E}}
\DeclareMathOperator{\Supp}{Supp}

\DeclareMathOperator{\Adv}{Adv}

\newcommand{\cF}{\mathcal{F}}
\newcommand{\cA}{\mathcal{A}}

\newcommand{\cD}{\mathcal{D}}

\newcommand{\Gen}{\mathrm{Gen}}
\newcommand{\Inv}{\mathrm{Inv}}
\newcommand{\Chk}{\mathrm{Chk}}
\newcommand{\Samp}{\mathrm{Samp}}
\newcommand{\td}{\mathsf{td}}
\newcommand{\Rk}{\mathcal{R}_k}

\newcommand{\hbin}{h}
\newcommand{\DeltaTr}[2]{\Delta\!\left(#1,#2\right)}
\newcommand{\nuG}{\nu_G}

\newcommand{\sC}{\mathsf{C}}
\newcommand{\sI}{\mathsf{I}}
\newcommand{\sB}{\mathsf{B}}
\newcommand{\Coin}{\Lambda}

\newcommand{\PsiY}{|\Psi_y\rangle}
\newcommand{\PiN}[1]{\tilde{\Pi}_{#1}^{(N)}}
\newcommand{\MN}[1]{\tilde{M}_{#1}^{(N)}}
\newcommand{\MBN}[1]{\overline{M}_{#1}^{(N)}}
\newcommand{\AdvN}{\Adv_N}

\title{From Claws to Conferences:\\ Device-Independent Key Agreement Beyond Non-Signaling}
\author{Suvradip Chakraborty\thanks{ORCID: \href{https://orcid.org/0000-0002-5352-4946}{0000-0002-5352-4946}.
  E-mail: \href{mailto:suvradip1111@gmail.com}{\texttt{suvradip1111@gmail.com}}}\\[2pt]
  {\normalsize Visa Research}}
\date{}

\begin{document}
\maketitle

\begin{abstract}
Device-independent conference key agreement (DI-CKA) allows a group of users to establish a common secret key
without trusting the internal operation of their quantum devices.  Existing DI-CKA protocols derive their security
from multipartite Bell tests and therefore require the device components not to communicate during each round.
We present, to the best of our knowledge, the first DI-CKA protocol that removes this non-signaling assumption:
all interfaces and their mutual communication are treated as one untrusted device, which only has to be quantum
polynomial-time while the protocol runs.  Assuming that the Learning With Errors (LWE) problem is hard for quantum
algorithms with quantum advice during the execution, we prove finite-size security against coherent attacks, in
which the device may keep quantum memory and correlate and adapt its behavior across rounds.  For two parties the
protocol is a computational DI-QKD protocol with an explicit finite-size proof against coherent attacks.

Technically, we lift the claw state underlying the classical-verifier test of Brakerski et al.\ (FOCS 2018) to
multipartite entanglement, obtaining a \emph{GHZ-correlated claw state}.
We then introduce the \emph{collective adaptive hardcore bit}, a multiparty generalization of the adaptive hardcore
bit of Brakerski et al.\ and show that it follows
from the original property by a tight reduction.  Combined with Jordan's lemma, an entropic uncertainty relation, and entropy accumulation, this certifies
smooth min-entropy of the key against an external quantum adversary.  Our protocol achieves \emph{everlasting security}:
LWE needs to hold only during the execution, because the final key is statistically secure and its secrecy persists
even if LWE is broken after the protocol ends.  
\end{abstract}

\section{Introduction}\label{sec:intro}

Quantum key distribution (QKD) allows two parties to establish a secret key whose confidentiality rests on the
laws of quantum mechanics rather than on the presumed hardness of a computational problem~\cite{BB84,Ekert1991}.
This guarantee, however, is only as reliable as the correspondence between the security proof and the physical
implementation.  Conventional proofs model sources, measurements, and detectors mathematically.  Real apparatus
deviates from these models, and such deviations have enabled side-channel attacks that break implementations
without contradicting quantum mechanics~\cite{Lydersen2010,Xu2020}.

\emph{Device independence} (DI) removes this modeling premise~\cite{MY1998,ABG2007,VV2014}.  The users retain
trusted classical control, private randomness, and authenticated classical communication, but treat their quantum
apparatus as an uncharacterized and possibly adversarial black box: security is inferred from the observed
classical input--output behavior alone, without any assumption on the states or measurements inside.  The strength
of this guarantee has motivated loophole-free Bell experiments~\cite{Hensen2015,Giustina2015,Shalm2015} and the
first experimental demonstrations of DI-QKD~\cite{Nadlinger2022,Zhang2022}.

In information-theoretic DI protocols the certificate is a \emph{Bell-inequality violation}: if the observed
correlations cannot be reproduced by any local hidden-variable model, then the device components must share
entanglement and perform incompatible measurements.  This inference requires that the components cannot
communicate while they are being measured.  This \emph{non-signaling assumption} is usually justified by
space-like separation, so that no signal travelling at or below the speed of light can pass between the
components before all measurements are completed.  It is not a technicality of the proof: it is the physical
mechanism from which the proof obtains its soundness.

\paragraph{The multiparty setting.}
Many applications require a key shared by a group rather than by a pair.  In \emph{conference key agreement}
(CKA), Alice and $N-1$ Bobs establish one common secret key, a basic primitive for secure group communication and
distributed protocols~\cite{MurtaReview2020}.  Existing DI-CKA protocols certify security through multipartite
Bell tests~\cite{Ribeiro2018,Holz2020} and therefore require non-signaling among all $N$ components during every
measurement.  Enforcing this is considerably harder than in the bipartite case.  The experiment must coordinate
$N$ laboratories, their quantum links, their random input choices, and tightly timed responses, and the number of
pairwise causal separations that must be maintained and audited grows quadratically with $N$.

\paragraph{Replacing non-signaling by computational hardness.}
Brakerski, Christiano, Mahadev, Vazirani, and Vidick (BCMVV)~\cite{BCMVV2018} showed that a different assumption
can take the place of non-signaling.  If the device is computationally bounded, a classical verifier can certify
its quantum behavior by interacting with a \emph{single} device, without any spatial separation.  Their test is
built from noisy trapdoor claw-free functions constructed from the Learning With Errors (LWE) problem and yields
both a test of quantumness and certifiable randomness.  Related techniques underlie the classical verification of
quantum computations~\cite{Mahadev2018Verification}, remote state preparation and blind delegation
\cite{GheorghiuVidick2019,BartusekKhurana2025}, and the compilation of nonlocal games into single-device
protocols~\cite{KLVY2023}.

For key distribution, Metger, Dulek, Coladangelo, and Arnon-Friedman~\cite{Metger2024} constructed a
computational DI-QKD protocol in which the two device components may exchange arbitrary quantum communication,
provided the device is computationally bounded while the protocol runs.  They proved security in the asymptotic
IID setting, in which the device behaves identically and independently in every round.  A \emph{coherent} attack
is more powerful: the device may keep quantum memory and correlate and adapt its behavior across all rounds.
Merkulov and Arnon~\cite{MA2025} subsequently developed a modular framework that combines entropic
uncertainty relations and entropy accumulation with post-quantum assumptions, and used it to certify randomness
from a single device against general adversaries.  To the best of our knowledge, however, no computational
approach to the multiparty task has been proposed, and every existing DI-CKA protocol relies on non-signaling
among the device components.  This leaves open the following question.
\begin{quote}
\emph{Can device-independent conference key agreement be achieved without any non-signaling assumption among the
device components, with security against coherent attacks?}
\end{quote}
We answer this question in the affirmative.

\subsection{Our Results}\label{sec:results}

\paragraph{Setting.}
Alice and $N-1$ Bobs have trusted classical control and authenticated public classical channels, and Alice holds
private randomness.  They share an untrusted quantum apparatus with one interface per participant.  For soundness,
all interfaces, their shared memory, and any classical or quantum communication between them are treated as a
\emph{single} quantum polynomial-time (QPT) device that may keep quantum memory and adapt to the entire past. In particular,  there
is no tensor-product, non-signaling, or space-like-separation assumption.  The eavesdropper Eve is computationally
unbounded, may hold a purification of the device's initial state, and sees all public communication.  As
in~\cite{Metger2024}, the device cannot send information to Eve during the execution; once the components may
communicate, this is unavoidable, as the device could otherwise simply forward the key.  The honest implementation
needs no communication between the components: pre-distributed GHZ states and local operations suffice
(Remark~\ref{rem:ghz-preshared}).  The protocol can therefore be layered on a standard DI setup in which physical
isolation is attempted but not relied upon~\cite{Metger2024}.

\begin{theorem}[Main result, informal]\label{thm:informal}
Let $N\geq2$, and assume that LWE is hard for QPT algorithms with polynomial-size quantum advice while the
protocol runs (Assumption~\ref{ass:ntcf}).  Then there is an $N$-party DI-CKA protocol (Protocol~\ref{prot:dicka}) with the following properties.
\begin{enumerate}
\item \emph{Security against coherent attacks.}  For every QPT device, even one whose components communicate
arbitrarily and which keeps quantum memory and adapts across rounds, the protocol is $\varepsilon_{\rm corr}$-correct
and $\varepsilon_{\rm sec}$-secret (Definition~\ref{def:dicka-security}): all Bobs output Alice's key, and this key
is close to uniform given Eve's information.  The key has length $nt-O(\sqrt n)-L$, where $n$ is the number of
rounds, $t$ is the certified min-entropy per round, and $L$ is the public leakage (Theorem~\ref{thm:main}).
\item \emph{Completeness.}  An honest noiseless implementation aborts only with probability exponentially small
in $n$, up to the failure probability of classical post-processing
(Proposition~\ref{prop:protocol-completeness}).
\item \emph{Everlasting security.}  The secrecy guarantee is statistical.  It remains valid against arbitrary
later computation, including an algorithm that breaks LWE after the protocol has ended
(Proposition~\ref{prop:lifting}).
\end{enumerate}
For $N=2$ the protocol is a computational DI-QKD protocol with the same guarantees (Corollary~\ref{cor:diqkd}).
\end{theorem}

\paragraph{Contributions.}
Theorem~\ref{thm:informal} rests on the following ingredients.
\begin{enumerate}
\item \textbf{GHZ-correlated claw state (Section~\ref{sec:mp-game}).}
We give a multiparty state-preparation procedure that combines GHZ-type entanglement with the claw structure of a 
trapdoor claw-free family. One coherent claw-free function evaluation followed by $N-1$ CNOT gates prepares the state
$(|0\rangle^{\otimes N}|x_0\rangle+|1\rangle^{\otimes N}|x_1\rangle)/\sqrt2$, which lifts the two-branch claw
structure of BCMVV to $N$ parties.  Its computational-basis measurement gives every participant the same bit,
and its Hadamard-basis measurement satisfies one global parity relation tied to the hidden claw.

\item \textbf{Collective adaptive hardcore bit (Section~\ref{sec:mp-ahb}).}
We formulate a multiparty generalization of the BCMVV adaptive hardcore bit (AHB) property~\cite{BCMVV2018}, tailored to the GHZ-correlated claw state, which we call the \emph{collective AHB} property.  Treating the whole apparatus, including its internal
communication, as one QPT circuit, we show that the collective property follows from the ordinary AHB property via a tight reduction that loses nothing, and in particular nothing that depends on $N$.  It therefore inherits the LWE-based
instantiation of BCMVV.

\item \textbf{DI-CKA secure against coherent attacks without non-signaling (Sections~\ref{sec:protocol}--\ref{sec:main}).}
Building on these two ingredients, we construct what is, to the best of our knowledge, the first DI-CKA protocol
whose security does not rely on non-signaling among the device components. We prove its finite-size
security against coherent attacks.  The proof adapts the single-device analysis of~\cite{MA2025} to key
agreement.  
For $N=2$, the protocol is a
computational DI-QKD protocol with an explicit finite-size security proof against coherent attacks.  

\item \textbf{Everlasting security (Section~\ref{sec:lifting}).}
The hardness of LWE is needed only during the challenge--response phase. Once privacy amplification is complete, the key is statistically close to uniform conditioned on the public transcript and Eve's side information. This guarantee is preserved under arbitrary subsequent quantum computation, including an algorithm that later solves the public LWE instances, provided that excluded device memory and erased verifier secrets remain inaccessible. Whereas earlier computational-DI works discuss this property informally, we provide an explicit trace-distance formulation and lifting theorem that makes the necessary isolation and erasure assumptions precise.

\item \textbf{Multiparty resource saving (Section~\ref{sec:discussion}).}
Distributing a conference key through $N-1$ pairwise computational DI-QKD sessions uses $N-1$ coherent
claw-free function evaluations per raw conference symbol.  Our construction  requires
only one such evaluation, together with $N-1$ CNOT gates.  This is a structural saving in the most expensive quantum circuit of the protocol, namely the state-preparation circuits.

\end{enumerate}

\paragraph{Scope and limitations.}
We assume that all conference participants are honest, and the adversary is external. We do not address insider attacks or
group-membership functionality.  As in~\cite{Metger2024}, the device's internal communication must be inaccessible to
Eve during the execution (Section~\ref{sec:model}).  Finally, our certified rate inherits the stringent noise
requirements of the single-round bound of~\cite{MA2025}: it vanishes once the aggregate test-failure probability
exceeds about $10^{-12}$, which corresponds to depolarizing noise of order $10^{-10}/(N-1)$ at best
(Table~\ref{tab:noise-threshold}).  Our results should therefore be read as a proof of principle, and any
improvement of the single-round bound translates directly into better rates for our protocol.

\subsection{Technical Overview}\label{sec:tech-overview}

\begin{figure}[t]
\centering
\resizebox{\textwidth}{!}{%
\begin{tikzpicture}[
 proofbox/.style={draw=black!65, rounded corners=2pt, align=center,
   font=\scriptsize, minimum width=2.7cm, minimum height=0.95cm, inner sep=2.5pt},
 source/.style={proofbox, fill=gray!12},
 comp/.style={proofbox, fill=blue!9},
 info/.style={proofbox, fill=green!9, dashed},
 result/.style={proofbox, fill=orange!14},
 flow/.style={->, >=latex, thick, draw=black!65}]
\node[source] (tcf)    at (0,1.35)   {TCF with AHB\\\cite{BCMVV2018}};
\node[comp]   (ghz)    at (3.2,1.35) {GHZ-correlated\\claw state (\S\ref{sec:mp-game})};
\node[comp]   (cahb)   at (6.4,1.35) {Collective AHB:\\tight reduction (\S\ref{sec:mp-ahb})};
\node[comp]   (bal)    at (9.6,1.35) {Balanced binary\\measurement (\S\ref{sec:mp-goodsub})};
\node[info]   (jordan) at (9.6,-0.2) {Jordan blocks and\\good subspace (\S\ref{sec:mp-goodsub})};
\node[info]   (eur)    at (6.4,-0.2) {Uncertainty relation:\\one-round entropy (\S\ref{sec:single-round})};
\node[info]   (eat)    at (3.2,-0.2) {Affine tradeoff\\and EAT (\S\ref{sec:main})};
\node[result] (key)    at (0,-0.2)   {Leakage and hashing:\\conference key (\S\ref{sec:main})};
\node[result] (ever)   at (0,-1.75)  {Everlasting\\security (\S\ref{sec:lifting})};
\draw[flow] (tcf)--(ghz);  \draw[flow] (ghz)--(cahb);  \draw[flow] (cahb)--(bal);
\draw[flow] (bal)--(jordan); \draw[flow] (jordan)--(eur); \draw[flow] (eur)--(eat);
\draw[flow] (eat)--(key);  \draw[flow] (key)--(ever);
\end{tikzpicture}}
\caption{Structure of the security proof.  The blue steps (solid frames, upper row) carry the computational
assumption, which enters only through the AHB property; the green steps (dashed frames) are information theoretic.  The final trace-distance guarantee
is lifted to everlasting security by data processing.}\label{fig:proof-architecture}
\end{figure}
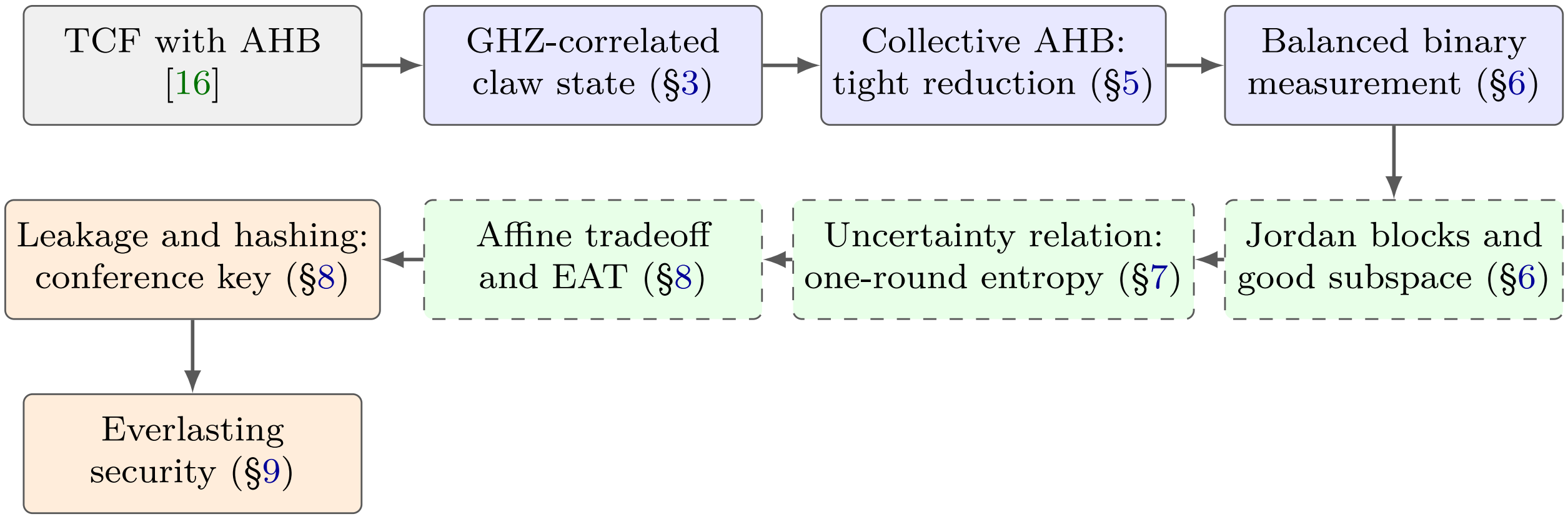

We now explain the main ideas.  Figure~\ref{fig:proof-architecture} shows the structure of the proof: the
computational assumption enters only through the upper row, and every step of the lower row is information
theoretic.

\paragraph{Starting point: a classically certified qubit.}
A trapdoor claw-free (TCF) family consists of pairs of injective functions $f_{k,0},f_{k,1}$ with a common image,
so that every image $y$ has a unique \emph{claw} $(x_0,x_1)$ with $f_{k,0}(x_0)=f_{k,1}(x_1)=y$.  The trapdoor of
the key $k$ reveals the claw, but without it no QPT algorithm can find one.  A quantum device can nevertheless
evaluate both branches in superposition and measure the image $y$, which leaves the state
\begin{equation}\label{eq:overview-claw-state}
 \tfrac{1}{\sqrt2}\bigl(|0\rangle|x_0\rangle+|1\rangle|x_1\rangle\bigr).
\end{equation}
BCMVV~\cite{BCMVV2018} pair this state with two complementary challenges.  In a \emph{preimage test}, the device
measures in the computational basis and returns a valid preimage $(b,x_b)$.  In an \emph{equation test}, it
measures in the Hadamard basis and returns a string $d$ and a bit $e$, which must satisfy
$e=d\cdot(x_0\oplus x_1)$.  The \emph{adaptive hardcore bit} (AHB) property states that no QPT device can do
both: even after returning a valid preimage, it cannot output an admissible $d$ together with a non-negligibly
biased prediction of $d\cdot(x_0\oplus x_1)$.  Admissibility means membership of $d$ in a set $G_{k,y}$ that
contains almost all strings and is recognizable with the trapdoor; it excludes degenerate choices such as
$d=0^w$, for which the equation is trivial.  This computational form of measurement incompatibility is what
lets a classical verifier certify quantum behavior.

\paragraph{Step 1: a conference correlation from a single claw.}
Key agreement needs more than certified randomness: Alice and every Bob must obtain the \emph{same} raw bit, and
the verifier must still be able to test the privacy of that bit.  We obtain both properties from one claw.
Starting from~\eqref{eq:overview-claw-state}, Alice fans the branch qubit out to $N-1$ fresh qubits with CNOT
gates, which produces the \emph{GHZ-correlated claw state}
\begin{equation}\label{eq:overview-ghz-claw}
 |\Psi_y\rangle=\tfrac{1}{\sqrt2}\bigl(|0\rangle^{\otimes N}|x_0\rangle+|1\rangle^{\otimes N}|x_1\rangle\bigr),
\end{equation}
and she sends the $j$-th branch qubit to Bob $j$.  A computational-basis measurement now gives every participant
the same branch bit $b$, a raw key symbol, and gives Alice the matching preimage $x_b$.  A Hadamard-basis
measurement of the preimage register and the branch registers gives outcomes $d$ and $u_1,\ldots,u_N$ that
satisfy the \emph{global parity relation} (Proposition~\ref{prop:xor})
\begin{equation}\label{eq:overview-parity}
 u_1\oplus\cdots\oplus u_N=d\cdot(x_0\oplus x_1).
\end{equation}
Thus a single coherent TCF evaluation and $N-1$ CNOT gates support both a common raw symbol and a test of its
privacy.  The state~\eqref{eq:overview-ghz-claw} describes only the honest implementation.  The soundness
analysis assumes nothing about the state or the measurements of the device.

\paragraph{Step 2: soundness for a communicating apparatus via collective AHB.}
The parity relation~\eqref{eq:overview-parity} involves answers produced by all $N$ interfaces, which may
communicate arbitrarily.  To constrain them, we consider an \emph{analytical} sequential experiment; the protocol
itself asks only one of the two questions in each round.  After the device has committed to $y$, apply its
preimage instrument and record Alice's answer $(b,x)$, leaving the Bobs' preimage answers unmeasured.  Then apply
its equation instrument to the residual state and record $(d,u_1,\ldots,u_N)$.  When Alice's preimage is valid, the equation answer falls into one of three
classes:
\begin{align*}
 \mathsf C&:\ d\in G_{k,y}\ \text{ and }\ {\textstyle\bigoplus_{j=1}^N}u_j=d\cdot(x_0\oplus x_1),\\
 \mathsf I&:\ d\in G_{k,y}\ \text{ and }\ {\textstyle\bigoplus_{j=1}^N}u_j\neq d\cdot(x_0\oplus x_1),\\
 \mathsf B&:\ d\notin G_{k,y}\ \text{ or the answer is malformed}.
\end{align*}
The \emph{collective AHB} property (Definition~\ref{def:mp-ahb}) asserts that
$|\Pr[\mathsf C]-\Pr[\mathsf I]|$ is negligible for every joint QPT device.  It follows from the ordinary AHB
property by an exact reduction (Theorem~\ref{thm:mp-ahb}).  Since all interfaces form one QPT machine, the AHB
adversary can simulate the entire apparatus, including its internal communication, and output the single
predicted bit $e=\bigoplus_ju_j$.  This map sends $\mathsf C$ and $\mathsf I$ exactly to the correct and
incorrect events of the ordinary AHB game.  Malformed answers are completed to well-formed ones with a uniformly
random final bit, so their contributions to the two events are equal and cancel.  Note that the events involve
only Alice's preimage answer.  Whether the Bobs' branch bits agree with Alice's is a correctness question, which
the key-agreement layer tests and repairs separately.

\paragraph{Step 3: the protocol.}
Protocol~\ref{prot:dicka} (Section~\ref{sec:protocol}) is run by a trusted classical verifier hosted by Alice, and
every round uses a fresh TCF key.  After the device has committed to an image, the verifier privately samples a
generation flag $G$ and reveals only a challenge $T$, where
\begin{equation}\label{eq:overview-round-law}
 \Pr[G=1]=1-\gamma,\qquad \Pr[T=0\mid G=1]=1,\qquad \Pr[T=1\mid G=0]=\beta.
\end{equation}
For $T=0$ the device returns a branch bit to every party and a preimage to Alice.  For $T=1$ it returns the
equation answers $(d,u_1,\ldots,u_N)$, which the verifier checks for admissibility and for the parity relation~\eqref{eq:overview-parity}. Only after the answer has been recorded does the verifier reveal $G$.
A $T=0$ answer is then either opened as a preimage test
($G=0$) or kept as a raw key symbol ($G=1$).  Because the device cannot distinguish these two cases, the preimage
tests constrain exactly the instrument that produces the key. The opened tests also estimate every Alice--Bob
disagreement rate.  All abort decisions are deferred until the fixed number $n$ of rounds is complete, so that the
execution is a fixed sequence of $n$ round channels.  This structure is what lets us use the entropy accumulation
theorem in Step~5 to turn a single-round entropy bound into one for the whole execution.  Finally, the Bobs reconcile their
raw strings with Alice's by one-way public communication and verify equality, and all parties apply a
two-universal hash function to obtain the conference key.

\paragraph{Step 4: from hardness to entropy in a single round.}
This is the technical core of the proof.  We must turn the computational statement of Step~2 into a lower bound
on the entropy of Alice's branch bit conditioned on Eve's quantum side information.  We follow the strategy
of~\cite{BCMVV2018,MA2025}, which consists of three information-theoretic steps.  The main new difficulty lies in
connecting the AHB property to the first of them.

\smallskip\noindent\emph{(a) A binary equation measurement with the right bias.}
The operator argument requires two binary projective measurements.  Its computational input is their
\emph{sequential bias}: after the preimage measurement, the outcome of the equation measurement should be close to
a fair coin.  The protocol accepts $\mathsf C$ and rejects $\mathsf I\cup\mathsf B$, but collective AHB controls
only $\Pr[\mathsf C]-\Pr[\mathsf I]$, and it says nothing about the weight of $\mathsf B$ after a preimage
measurement.  We therefore introduce a \emph{balanced completion}.  We adjoin an independent fair coin $R$ and
split the class $\mathsf B$ evenly between the two outcomes:
\begin{equation}\label{eq:overview-balanced-measurement}
 \overline M_0=M_{\mathsf C}\otimes\id_R+M_{\mathsf B}\otimes|0\rangle\!\langle0|_R,\qquad
 \overline M_1=M_{\mathsf I}\otimes\id_R+M_{\mathsf B}\otimes|1\rangle\!\langle1|_R.
\end{equation}
Then $\Pr[\overline M_0]-\Pr[\overline M_1]=\Pr[\mathsf C]-\Pr[\mathsf I]$, so the sequential bias of
$\{\overline M_0,\overline M_1\}$ equals the collective AHB advantage.  Moreover
$\Pr[\overline M_0]\geq\Pr[\mathsf C]$, so the success probability of $\overline M_0$ is at least the observed
test success (Lemma~\ref{lem:balanced-completion}).  The completion is used only in the analysis; the protocol
still rejects every inadmissible $d$.

\smallskip\noindent\emph{(b) Most of the state lies in complementary blocks.}
Jordan's lemma (Lemma~\ref{lem:jordan}) decomposes the underlying space into blocks of dimension at most two that
are invariant under both Alice's preimage projector, restricted to valid preimages, and $\overline M_0$.  Within
each block the two measurements are related by a single angle, and their overlap $c\in[1/2,1]$ quantifies how far
they are from complementary.  The good-subspace lemma (Lemma~\ref{lem:goodsub}) shows that a small sequential
bias and a high equation success together force almost all of the weight of the state onto blocks of small
overlap.  Jordan's lemma applies to any pair of projections, so the blocks need not respect parties,
laboratories, or tensor factors.  This is precisely why the operator argument survives arbitrary communication
among the interfaces.

\smallskip\noindent\emph{(c) Uncertainty.}
On the good blocks, an entropic uncertainty relation~\cite{TomamichelRenner2011} converts small overlap and high
predictability of the equation outcome into entropy of Alice's branch bit conditioned on Eve.  Accounting for the
gentle projection onto valid preimages, the weight of the bad blocks, and a continuity bound gives the one-round
bound (Theorem~\ref{thm:single-round})
\begin{equation}\label{eq:overview-one-round-invariant}
 H(\widehat\Pi\mid EYK)\geq g(\omega_p,\omega_m)-\xi_{\rm AHB}(\lambda),
\end{equation}
where $\widehat\Pi\in\{0,1,\perp\}$ is Alice's coarse preimage outcome, $\omega_p$ and $\omega_m$ are the success
probabilities of the two tests, $g$ is an explicit function, and $\xi_{\rm AHB}$ is negligible.  The number of
parties affects the success probabilities but not the function $g$.

\paragraph{Step 5: from one round to coherent attacks.}
The entropy accumulation theorem (EAT)~\cite{DFR2020,MFSR2022} converts a per-round bound that is affine in the
observed statistics into a smooth min-entropy bound for a sequential device with arbitrary memory.  Two points
require care.  First, the per-round bound must hold for every state that the device may carry into a round.
Such a state may depend on the entire past and need not be efficiently preparable.  Because each round uses a
fresh key, however, the carried state is independent of that key and has polynomial size.  It is therefore a
legitimate quantum advice state for the AHB adversary~\cite{MA2025}, and~\eqref{eq:overview-one-round-invariant}
applies after every history.  Second, the protocol observes only an aggregate test score, not the pair
$(\omega_p,\omega_m)$.  We minimize $g$ over all pairs compatible with the score and support a convex minorant of
the result by an affine function, which gives the min-tradeoff function (Lemma~\ref{lem:min-tradeoff}).  The EAT
then certifies the complete coarse-grained output.  Removing the test symbols and charging every public
message---opened tests, error-correction syndrome, and verification tags---by its support size leaves an
extractable key of length
\begin{equation}\label{eq:overview-key-budget}
 \ell\le nt-v\sqrt n-L_{\rm drop}-L_{\rm test}-L_{\rm EC}-2\log\frac1{2\varepsilon_{\rm PA}},
\end{equation}
where $t$ is the certified per-round entropy and $v\sqrt n$ is the finite-size correction of the EAT.  The
leakage terms $L_{\rm drop}$ and $L_{\rm test}$ pay for removing the test symbols and for the publicly announced
test data, $L_{\rm EC}$ pays for the error-correction syndrome and the verification tags, and the last term is the
cost of privacy amplification with error $\varepsilon_{\rm PA}$ (Theorem~\ref{thm:main} gives all terms explicitly).  The
carried register contains all interfaces and their memory, so no IID or non-signaling assumption enters at any
point.

\paragraph{Step 6: everlasting security.}
The final guarantee is a trace-distance bound.  The real state $\rho_{K_ACE}$ of Alice's key $K_A$, the public
transcript $C$ and Eve's quantum system $E$ is $\varepsilon$-close to the ideal state $\tau_{K_A}\otimes\rho_{CE}$, in which the key is
uniform and independent of $CE$.  LWE is used to prove this bound, but the bound itself is statistical.
Whatever Eve does after the protocol is some quantum channel $\Phi$ on $CE$, of arbitrary complexity.  By the
data-processing inequality, $\mathrm{id}_{K_A}\otimes\Phi$ cannot increase the trace distance, and it maps the
ideal state to $\tau_{K_A}\otimes\Phi(\rho_{CE})$, which is again ideal.  The key therefore stays
$\varepsilon$-close to uniform given anything Eve computes later, even with an algorithm that breaks the LWE
instances in the transcript.  The computational assumption is thus needed only while the device answers the
challenges.
Proposition~\ref{prop:lifting} and Appendix~\ref{app:lifting} make this precise, including the isolation and erasure
conditions on which it relies.

\paragraph{Relation to~\cite{MA2025}.}
Merkulov and Arnon~\cite{MA2025} developed the single-device toolkit on which our analysis builds: the
combination of Jordan's lemma, the good-subspace lemma, and entropic uncertainty with entropy accumulation, as a
modular alternative to the original multi-round analysis of BCMVV.  Our single-round function $g$ has the same
form as theirs~\cite[Lemma~31]{MA2025}.  Using this toolkit for key agreement requires components that a
randomness-certification protocol does not provide: the GHZ-correlated claw state and its global parity test, the
collective AHB property, an agreement test, and the reconciliation, verification, and leakage accounting needed
for a common key that is secret from an external eavesdropper.  In addition, the binary equation measurement
of~\cite{MA2025} maps every inadmissible answer to rejection, whereas the AHB property controls only
$\Pr[\mathsf C]-\Pr[\mathsf I]$.  The balanced completion of Step~4(a) makes the sequential bias of the binary
measurement exactly equal to the AHB advantage.

\subsection{Related Work}\label{sec:related}

\paragraph{DI key distribution from nonlocality.}
DI-QKD was proposed in~\cite{MY1998,ABG2007} and proved secure against general attacks in~\cite{VV2014}; entropy
accumulation~\cite{DFR2020,MFSR2022} now underlies most finite-size analyses~\cite{AFDFRV2018}.  In the multiparty setting, Ribeiro,
Murta, and Wehner introduced fully device-independent CKA~\cite{Ribeiro2018}, and later work certified the same
task with other multipartite Bell inequalities~\cite{Holz2020}; see~\cite{MurtaReview2020} for a review.  Further
quantum CKA schemes, such as those based on single-photon interference~\cite{Grasselli2019}, do not by themselves
provide a fully device-independent guarantee.  All of these protocols derive their soundness from non-signaling
among the device components.  Our model instead allows arbitrary communication inside a QPT device and relies on a
cryptographic assumption during the execution.  The two are different trust models, neither of which implies the
other.

\paragraph{Computational device independence.}
BCMVV~\cite{BCMVV2018} introduced noisy trapdoor claw-free functions with the adaptive hardcore bit property and
used them for a test of quantumness and for certified randomness; Mahadev used related machinery to verify
quantum computations classically~\cite{Mahadev2018Verification}.  Subsequent work developed remote and oblivious
state preparation and blind delegation~\cite{GheorghiuVidick2019,BartusekKhurana2025}.  Kalai, Lombardi,
Vaikuntanathan, and Yang~\cite{KLVY2023} showed how to compile any nonlocal game into a single-device protocol,
and quantitative soundness has recently been established for all multipartite compiled games~\cite{Baroni2025}.
Compiled multipartite Bell tests may offer an alternative route to computational DI-CKA, which we do not pursue
here.  Certified randomness has also been obtained from random circuit sampling under different complexity
assumptions~\cite{AH2023}.  The works closest to ours are the computational DI-QKD protocol of
Metger et al.~\cite{Metger2024} and the single-device entropy accumulation framework of Merkulov and
Arnon~\cite{MA2025}, both discussed above.

\paragraph{Post-quantum group key exchange.}
A post-quantum group key exchange, such as the Ring-LWE protocol of Apon et al.~\cite{Apon2019}, solves a
different problem.  Its participants run specified classical algorithms, and the session key remains secure only
as long as the underlying problem remains hard.  It does not certify untrusted quantum apparatus.  Conversely, our
protocol assumes authenticated classical channels and does not provide the authentication or membership
functionality of a group key exchange.  Its contribution is to turn observations of an untrusted, internally
communicating quantum device into a common secret key with a statistical, and hence everlasting, security
guarantee (Section~\ref{sec:pq-gke-comparison}).

\subsection{Organization}\label{sec:organization}

Section~\ref{sec:prelim} fixes notation, the security definition, and the cryptographic and information-theoretic
tools.  Section~\ref{sec:mp-game} introduces the GHZ-correlated claw state and the observables of an arbitrary
device, and Section~\ref{sec:protocol} presents the protocol and states the main theorem.
Sections~\ref{sec:mp-ahb}--\ref{sec:main} prove it: collective AHB (Section~\ref{sec:mp-ahb}), the balanced
binary measurement and the good-subspace bound (Section~\ref{sec:mp-goodsub}), the single-round entropy bound
(Section~\ref{sec:single-round}), and multi-round composition (Section~\ref{sec:main}).
Section~\ref{sec:lifting} establishes everlasting security, and Section~\ref{sec:discussion} discusses honest
noise and resource costs.  Full derivations appear in
Appendices~\ref{app:collective-ahb}--\ref{app:lifting}.

\section{Preliminaries and Security Definition}\label{sec:prelim}

All logarithms are base two.  For normalized or subnormalized positive operators, write
\begin{equation}\label{eq:trace-distance}
 \DeltaTr{\rho}{\sigma}:=\tfrac12\|\rho-\sigma\|_1.
\end{equation}
Trace distance is contractive under quantum channels.  We write $H(A|B)$ for the conditional von Neumann entropy
and $H_{\min}^{\varepsilon}(A|B)$ for the smooth conditional min-entropy.  If $L$ is a classical register with
support of size at most $2^s$, then~\cite{Renner2008}
\begin{equation}\label{eq:classical-leakage}
 H_{\min}^{\varepsilon}(Z|EL)\geq H_{\min}^{\varepsilon}(Z|E)-s,
 \qquad
 H_{\min}^{\varepsilon}(Z|EL)\geq H_{\min}^{\varepsilon}(ZL|E)-s.
\end{equation}

\subsection{Security definition}\label{sec:security-def}

A DI-CKA protocol is run by Alice $P_1$ and Bobs $P_2,\ldots,P_N$.  Let $\Omega$ be the event that it does not
abort and $p_\Omega=\Pr[\Omega]$.  On $\Omega$, Alice outputs a key $K_A\in\{0,1\}^\ell$ and each Bob $P_j$ outputs a
key $K_j\in\{0,1\}^\ell$; on abort, all parties output the symbol $\perp$.  Let $C$ denote the public transcript and
$E$ Eve's quantum system at the end of the protocol.

\begin{definition}[Correctness, secrecy, and completeness]\label{def:dicka-security}
The protocol is $\varepsilon_{\rm corr}$-correct if
\begin{equation}\label{eq:correctness-definition}
 \Pr[\Omega\wedge(\exists j\geq2:K_j\neq K_A)]\leq\varepsilon_{\rm corr}.
\end{equation}
It is $\varepsilon_{\rm sec}$-secret if
\begin{equation}\label{eq:secrecy-definition}
 p_\Omega\DeltaTr{\rho_{K_ACE|\Omega}}{\tau_{K_A}\otimes\rho_{CE|\Omega}}
 \leq\varepsilon_{\rm sec},
\end{equation}
where $\tau_{K_A}$ is the uniform distribution on $\{0,1\}^\ell$.  It is $\varepsilon_{\rm comp}$-complete for an honest
implementation if $\Pr[\neg\Omega]\leq\varepsilon_{\rm comp}$.
\end{definition}

The factor $p_\Omega$ prevents a device that accepts only on a rare, completely insecure branch from being
judged secure merely because the normalized conditional state is considered without its probability.
Equivalently, one can write the definition directly for the subnormalized accepting states.  Correctness and
secrecy compose with additive errors~\cite{BenOr2005,Canetti2001}.

\subsection{Noisy trapdoor claw-free functions}\label{sec:ntcf}

We use the noisy trapdoor claw-free (NTCF) families of BCMVV~\cite{BCMVV2018}, in the form stated
in~\cite{MA2025}.  For a finite set $\mathcal Y$, let $D_{\mathcal Y}$ be the set of probability distributions on
$\mathcal Y$, and for $p,q\in D_{\mathcal Y}$ let $H^2(p,q)=1-\sum_{y\in\mathcal Y}\sqrt{p(y)q(y)}$ be their squared Hellinger distance.  BCMVV allow an
efficient injective encoding of the domain into $\{0,1\}^w$. To simplify notation we identify the domain with its
encoding, so that $\mathcal X\subseteq\{0,1\}^w$ for some $w=\poly(\lambda)$.

\begin{definition}[NTCF family~\cite{BCMVV2018}]\label{def:ntcf}
Let $\mathcal X\subseteq\{0,1\}^w$, $\mathcal Y$, and $\mathcal K$ be finite sets.  A family
$\cF=\{f_{k,b}:\mathcal X\to D_{\mathcal Y}\}_{k\in\mathcal K,b\in\{0,1\}}$ is an NTCF family if the following
hold.
\begin{enumerate}
\item \emph{Key generation.}  A PPT algorithm $\Gen(1^\lambda)$ outputs a key $k\in\mathcal K$ and a trapdoor
$\td$.
\item \emph{Trapdoor injective pair.}  For all $k$, $b$, and $x\neq x'$, the supports of $f_{k,b}(x)$ and
$f_{k,b}(x')$ are disjoint, and an efficient deterministic algorithm satisfies $\Inv(\td,b,y)=x$ for all
$y\in\Supp(f_{k,b}(x))$.  Moreover, there is a perfect matching $\Rk\subseteq\mathcal X\times\mathcal X$ such that
$f_{k,0}(x_0)=f_{k,1}(x_1)$ if and only if $(x_0,x_1)\in\Rk$.  Its elements are called \emph{claws}.
\item \emph{Efficient range superposition.}  For all $k$ and $b$ there is a function
$f'_{k,b}:\mathcal X\to D_{\mathcal Y}$ such that:
(a) for all $(x_0,x_1)\in\Rk$ and $y\in\Supp(f'_{k,b}(x_b))$, $\Inv(\td,b,y)=x_b$ and
$\Inv(\td,1-b,y)=x_{1-b}$;
(b) an efficient deterministic algorithm $\Chk$, which is not given the trapdoor, outputs $\Chk(k,b,x,y)=1$ if
and only if $y\in\Supp(f'_{k,b}(x))$;
(c) $\E_{x\leftarrow\mathcal X}H^2(f_{k,b}(x),f'_{k,b}(x))\leq\eta_H(\lambda)$ for a negligible $\eta_H$, and an
efficient circuit $\Samp_{k,b}$ prepares
$|\mathcal X|^{-1/2}\sum_{x\in\mathcal X,y\in\mathcal Y}\sqrt{f'_{k,b}(x)(y)}\,|x\rangle|y\rangle$.
\item \emph{Admissible equation strings.}  For all $k$, $b$, and $x$ there is a set
$G_{k,b,x}\subseteq\{0,1\}^w$ such that $\Pr_{d\leftarrow\{0,1\}^w}[d\notin G_{k,b,x}]\leq\eta_G(\lambda)$ for a
negligible $\eta_G$, and membership in $G_{k,b,x}$ can be decided efficiently given $(k,b,x,\td)$.
\end{enumerate}
\end{definition}

\begin{definition}[Adaptive hardcore bit~\cite{BCMVV2018,MA2025}]\label{def:ahb-2party}
Let
\begin{align*}
 H_k&=\{(b,x_b,d,d\cdot(x_0\oplus x_1)):b\in\{0,1\},\ (x_0,x_1)\in\Rk,\
 d\in G_{k,0,x_0}\cap G_{k,1,x_1}\},\\
 \bar H_k&=\{(b,x,d,e):(b,x,d,1\oplus e)\in H_k\}.
\end{align*}
The family has the \emph{adaptive hardcore bit} (AHB) property if for every QPT algorithm $\cA$ there is a
negligible function $\eta_{\rm AHB}$ such that, for every polynomial-size quantum advice state $\phi$ that is
independent of $k$ but possibly inefficient to prepare,
\begin{equation}\label{eq:ordinary-ahb}
 \big|\Pr[\cA(k,\phi)\in H_k]-\Pr[\cA(k,\phi)\in\bar H_k]\big|
 \leq\eta_{\rm AHB}(\lambda),
\end{equation}
where the probabilities are over $(k,\td)\leftarrow\Gen(1^\lambda)$ and the measurements of $\cA$.
\end{definition}

A single function $\eta_{\rm AHB}$ can serve all advice states: for each $\lambda$, the advantage is a continuous
function on the compact set of advice states of the given size, and the maximizing states form an advice family
to which the hypothesis applies.

\begin{assumption}\label{ass:ntcf}
There exists an NTCF family (Definition~\ref{def:ntcf}) with the AHB property (Definition~\ref{def:ahb-2party}).
\end{assumption}

BCMVV construct an NTCF family from LWE~\cite[Section~4]{BCMVV2018}.  As observed in~\cite{MA2025}, this
construction satisfies Definition~\ref{def:ahb-2party} provided LWE is hard for QPT algorithms with
polynomial-size quantum advice.  Advice is needed in our setting because conditioning on earlier rounds can leave
the device in a polynomial-size state that is not efficiently preparable; fresh keys ensure that this state is
independent of the next key (Lemma~\ref{lem:fresh-key-history}).  The admissibility condition in $H_k$ is
essential: without it, $d=0^w$ would make the alleged hardcore bit identically zero.

The verifier never learns the claw directly; she recovers it from the image with the trapdoor.  The following
lemma shows that this is consistent with every valid preimage the device can produce.

\begin{lemma}[Well-formed images]\label{lem:well-formed}
Fix $(k,\td)$ and $y\in\mathcal Y$, and put $x_b(y):=\Inv(\td,b,y)$ for $b\in\{0,1\}$.  Call $y$ \emph{well formed}
if $\Chk(k,b,x_b(y),y)=1$ for some $b$.  If some $(b,x)$ satisfies $\Chk(k,b,x,y)=1$, then $y$ is well formed and
$x=x_b(y)$.  If $y$ is well formed, then $(x_0(y),x_1(y))\in\Rk$.
\end{lemma}

\begin{proof}
Suppose $\Chk(k,b,x,y)=1$, that is, $y\in\Supp(f'_{k,b}(x))$.  Since $\Rk$ is a perfect matching, there is a claw
$(x_0,x_1)\in\Rk$ with $x_b=x$.  Property~3(a) of Definition~\ref{def:ntcf} gives $\Inv(\td,b,y)=x$ and
$\Inv(\td,1-b,y)=x_{1-b}$.  Hence $x=x_b(y)$, the image $y$ is well formed, and $(x_0(y),x_1(y))=(x_0,x_1)\in\Rk$.
If $y$ is well formed, the same argument applied to $(b,x_b(y))$ proves the last claim.
\end{proof}

For a well-formed image we call $(x_0(y),x_1(y))$ its \emph{claw}, abbreviate it as $(x_0,x_1)$ when $y$ is clear,
and write
\begin{equation}\label{eq:good-set}
 G_{k,y}:=G_{k,0,x_0(y)}\cap G_{k,1,x_1(y)},\qquad \Pr_{d}[d\notin G_{k,y}]\leq2\eta_G(\lambda).
\end{equation}
By Lemma~\ref{lem:well-formed}, an image that is not well formed admits no valid preimage at all.

\paragraph{Noiseless special case.}
If every $f_{k,b}(x)$ is a point distribution and $f'_{k,b}=f_{k,b}$, then $f_{k,0}$ and $f_{k,1}$ are injective
functions with a common image, every image in that range is well formed, and $\eta_H=0$.  We describe the honest
state in this special case in Section~\ref{sec:state-prep} and treat general families in
Remark~\ref{rem:ntcf-honest}.

\subsection{Information-theoretic tools}\label{sec:it-tools}

For POVMs $\mathsf P=\{P_z\}$ and $\mathsf M=\{M_x\}$, define their square overlap by
\begin{equation}
	c(\mathsf P,\mathsf M)=\max_{z,x}\left\|\sqrt{P_z}\sqrt{M_x}\right\|_\infty^2.
\end{equation}
For every state on $AE\tilde{E}$ and measurements $\mathsf P,\mathsf M$ on $A$, the tripartite entropic uncertainty
relation~\cite{TomamichelRenner2011} gives
\begin{equation}\label{eq:eur}
	H(\mathsf P|E)+H(\mathsf M|\tilde{E})\ge \log(1/c(\mathsf P,\mathsf M)).
\end{equation}

\begin{lemma}[Jordan's lemma]\label{lem:jordan}
	Let $P$ and $M$ be Hermitian projections on a finite- or countable-dimensional Hilbert space.  There is an
	orthogonal decomposition into subspaces of dimension at most two that are invariant under both projections.
	Any normal operator commuting with both projections can be chosen block diagonal in the same decomposition.
\end{lemma}

On a two-dimensional block, an appropriate basis gives
\begin{equation}\label{eq:jordan-canonical}
	P=\begin{pmatrix}1&0\\0&0\end{pmatrix},\qquad
	M=\begin{pmatrix}
		\cos^2\theta&\cos\theta\sin\theta\\
		\cos\theta\sin\theta&\sin^2\theta
	\end{pmatrix}.
\end{equation}
The relevant overlap is $\max\{\cos^2\theta,\sin^2\theta\}\in[1/2,1]$.  The lemma concerns two global
projections; it makes no assertion about, and needs no factorization of, the underlying device.

\begin{lemma}[Gentle projection~\cite{Winter1999,Ogawa2005}]\label{lem:gentle}
If $P$ is a projection and $\rho$ is a state with $\Tr(P\rho)\geq1-\delta$, then
\begin{equation}
	\left\|\rho-\frac{P\rho P}{\Tr(P\rho)}\right\|_1\le2\sqrt\delta,
	\qquad
	\left\|\rho-P\rho P\right\|_1\le2\sqrt\delta.
\end{equation}
\end{lemma}

\begin{lemma}[Conditional-entropy continuity~\cite{AlickiFannes2004,Winter2016}]\label{lem:afw}
If $A$ is classical with alphabet size $d_A$ and
$\frac12\|\rho_{AB}-\sigma_{AB}\|_1\leq\epsilon\leq1$, then
\begin{equation}\label{eq:afw}
 |H(A|B)_\rho-H(A|B)_\sigma|
 \leq\epsilon\log d_A+(1+\epsilon)\hbin\!\left(\frac{\epsilon}{1+\epsilon}\right).
\end{equation}
\end{lemma}

\subsection{Entropy accumulation}\label{sec:eat-prelim}

We use the entropy accumulation theorem (EAT) of~\cite{DFR2020,MFSR2022}.  A round channel maps the carried
device memory $R_{i-1}$ to a new memory $R_i$, a finite classical output $O_i$, classical side information $S_i$,
and a finite classical score $Q_i$.  The score is a deterministic function of the output and the side
information, and the final state satisfies the Markov conditions
\begin{equation}\label{eq:eat-markov-prelim}
 O^{i-1}\ \longleftrightarrow\ S^{i-1}E\ \longleftrightarrow\ S_i,\qquad i=1,\ldots,n.
\end{equation}
An affine function $f_{\min}$ on score distributions is a min-tradeoff function if, for every round and every
input state that induces the score distribution $q$,
\begin{equation}\label{eq:min-tradeoff-definition}
	H(O_i|S_iR')\ge f_{\min}(q),
\end{equation}
where $R'$ is an arbitrary purifying reference.  Let $\Omega_0$ be an event on which the empirical score
distribution lies in a convex set $\widehat\Omega_0$, and let $t=\inf_{q\in\widehat\Omega_0}f_{\min}(q)$.  Then
\begin{equation}\label{eq:eat-symbolic}
	H_{\min}^{\varepsilon_s}(O^n|S^nE)_{\rho|\Omega_0}
	\ge nt-v(f_{\min},d_O,\varepsilon_s,p_{\Omega_0})\sqrt n,
\end{equation}
where $d_O$ is the one-round output dimension and $v$ is the explicit second-order constant of the chosen EAT
statement; for instance,~\cite{DFR2020} gives
\begin{equation}\label{eq:eat-v}
 v=2\bigl(\log(1+2d_O)+\lceil\|\nabla f_{\min}\|_\infty\rceil\bigr)\sqrt{1-2\log(\varepsilon_sp_{\Omega_0})}.
\end{equation}
Two features of $v$ matter below.  First, it grows with the slope of $f_{\min}$, so a nonconstant
tradeoff curve may not be replaced by a constant evaluated only at the acceptance threshold.  Second, it is
nonincreasing in $p_{\Omega_0}$, which depends on the device and is unknown.  We therefore fix
$\varepsilon_{\rm acc}\in(0,1]$ in advance and use $v:=v(f_{\min},d_O,\varepsilon_s,\varepsilon_{\rm acc})$;
\eqref{eq:eat-symbolic} then holds whenever $p_{\Omega_0}\geq\varepsilon_{\rm acc}$, and the complementary case
is handled in the security proof.

\section{Multiparty State Preparation and Measurements}\label{sec:mp-game}

We first describe the state and measurements of the honest implementation and then define the coarse-grained
projectors induced by an arbitrary joint device.  Only the latter enter the soundness proof.

\subsection{The GHZ-correlated claw state}\label{sec:state-prep}

Fix a key $k$ and a well-formed image $y$ with claw $(x_0,x_1)$.  The target state on $N$ branch registers and
Alice's preimage register $X$ is
\begin{equation}\label{eq:target-state}
	\PsiY=
	\frac{|0\rangle_{A_1}\cdots|0\rangle_{A_N}|x_0\rangle_X
		+|1\rangle_{A_1}\cdots|1\rangle_{A_N}|x_1\rangle_X}{\sqrt2}.
\end{equation}
For clarity we describe its preparation for a noiseless family; Remark~\ref{rem:ntcf-honest} treats general NTCF
families.  Applying $\Samp_{k,b}$ controlled on a branch qubit in the state $|+\rangle$ prepares the range
superposition
\begin{equation}\label{eq:range-superposition}
 \frac1{\sqrt{2|\mathcal X|}}\sum_{b\in\{0,1\},x\in\mathcal X}|b\rangle|x\rangle|f_{k,b}(x)\rangle.
\end{equation}

\begin{protocolbox}[title={Multiparty State-Preparation Protocol}]
	\textbf{Input.} A public key $k$ for the NTCF family.

	\smallskip
	\textbf{Registers.} Alice holds a branch qubit $A_1$, a preimage register $X$, and an image register $Y$.
	She also initializes auxiliary branch qubits $A_2,\ldots,A_N$ in $|0\rangle^{\otimes(N-1)}$.

	\smallskip
	\textbf{Output.} A classical image $y$ and the state~\eqref{eq:target-state}, with branch register $A_j$
	delivered to party $P_j$.

	\smallskip
	\textbf{Procedure.}
	\begin{enumerate}
		\item \textbf{Prepare one claw state.} On $(A_1,X,Y)$, Alice prepares the range superposition
		\eqref{eq:range-superposition}.

		\item \textbf{Collapse to a claw.} Alice measures $Y$ in the computational basis and obtains $y$.
		The remaining state on $(A_1,X)$ is
		\begin{equation*}
			\frac{|0\rangle_{A_1}|x_0\rangle_X+|1\rangle_{A_1}|x_1\rangle_X}{\sqrt2}.
		\end{equation*}

		\item \textbf{Create the GHZ correlation.} For every $j=2,\ldots,N$, Alice applies
		$\mathrm{CNOT}_{A_1\to A_j}$.  The resulting state is exactly $\PsiY$.

		\item \textbf{Distribute.} Alice keeps $(A_1,X)$ and sends $A_j$ to party $P_j$ for every
		$j=2,\ldots,N$.
	\end{enumerate}
\end{protocolbox}

\noindent Thus the multiparty layer adds only $N-1$ CNOT gates to one coherent NTCF evaluation.

\begin{proposition}[Exact honest preparation]\label{prop:exact-preparation}
For a noiseless family, measuring the image register of~\eqref{eq:range-superposition} and applying the CNOT
fan-out prepares exactly the state~\eqref{eq:target-state}, where $(x_0,x_1)$ is the claw of the observed image.
Every image in the common range occurs with probability $1/|\mathcal X|$.
\end{proposition}

\begin{proof}
	Injectivity of each branch and equality of their images imply that exactly two summands
	in~\eqref{eq:range-superposition} have image $y$, namely $(0,x_0)$ and $(1,x_1)$.  Their amplitudes are equal, so
	the image has probability $2/(2|\mathcal X|)=1/|\mathcal X|$ and the normalized post-measurement state is
	$(|0\rangle_{A_1}|x_0\rangle_X+|1\rangle_{A_1}|x_1\rangle_X)/\sqrt2$.  For $b\in\{0,1\}$, the CNOT fan-out maps
	$|b\rangle_{A_1}|0\rangle^{\otimes N-1}$ to $|b\rangle^{\otimes N}$.  Applying this unitary to both terms
	gives~\eqref{eq:target-state} without approximation error.
\end{proof}

\noindent Computational-basis measurement of the branch registers gives the same bit $b$ to all parties and
leaves Alice with the matching preimage $x_b$.  The Hadamard-basis behavior is captured by the following exact
parity identity.

\begin{proposition}[Multiparty equation]\label{prop:xor}
Measure every branch register and the preimage register of~\eqref{eq:target-state} in the Hadamard basis.  If
their outcomes are $u_1,\ldots,u_N$ and $d$, respectively, then every outcome of nonzero probability satisfies
\begin{equation}\label{eq:xor-condition}
 \bigoplus_{j=1}^Nu_j=d\cdot(x_0\oplus x_1).
\end{equation}
Moreover, $d$ is uniform and, conditioned on $d$, the $N$-bit outcome is uniform over the strings satisfying this
parity.
\end{proposition}

\begin{proof}
	For $b\in\{0,1\}$,
	\begin{align}
		H^{\otimes N}|b\rangle^{\otimes N}
		&=2^{-N/2}\sum_{\vec u\in\{0,1\}^N}
		(-1)^{b(\oplus_j u_j)}|\vec u\rangle,\\
		H^{\otimes w}|x_b\rangle
		&=2^{-w/2}\sum_{d\in\{0,1\}^w}(-1)^{d\cdot x_b}|d\rangle.
	\end{align}
	Consequently the amplitude of $(\vec u,d)$ is
	\begin{equation}\label{eq:joint-amplitude}
		\frac{(-1)^{d\cdot x_0}}{\sqrt{2^{N+w+1}}}
		\left[1+(-1)^{\oplus_j u_j+d\cdot(x_0\oplus x_1)}\right].
	\end{equation}
	The bracket vanishes unless~\eqref{eq:xor-condition} holds.  When it holds, the bracket has magnitude two, so
	every allowed pair $(\vec u,d)$ has probability $2^{1-N-w}$.  For each fixed $d$, exactly $2^{N-1}$ strings
	$\vec u$ have the required parity.  Summing their probabilities gives $2^{-w}$, which proves that $d$ is
	uniform; dividing by this marginal shows that $\vec u$ is uniform over the parity class.
\end{proof}

\begin{remark}[Preparation from pre-shared GHZ states]\label{rem:ghz-preshared}
The honest implementation needs no communication between the device components during a round.  Suppose the
parties pre-share $(|0\rangle^{\otimes N}+|1\rangle^{\otimes N})/\sqrt2$ on $A_1\cdots A_N$, and Alice applies
$\Samp_{k,b}$ to $(X,Y)$ controlled on her share $A_1$.  This produces
$(2|\mathcal X|)^{-1/2}\sum_{b,x,y}\sqrt{f'_{k,b}(x)(y)}\,|b\rangle^{\otimes N}|x\rangle_X|y\rangle_Y$, which is the
state obtained in the protocol above after the CNOT fan-out but before the image is measured.  Hence an honest
device succeeds with pre-distributed entanglement and local operations alone, as in the standard non-signaling
setting.  The communication permitted by our model is needed only to describe dishonest devices, in line with
the requirement of~\cite{Metger2024}.
\end{remark}

\begin{remark}[General NTCF families]\label{rem:ntcf-honest}
For a general family, Alice uses $\Samp_{k,b}$ to prepare
\begin{equation}
 |\phi'\rangle=\frac{1}{\sqrt{2|\mathcal X|}}\sum_{b,x,y}\sqrt{f'_{k,b}(x)(y)}\,|b\rangle|x\rangle|y\rangle.
\end{equation}
Let $|\phi\rangle$ be the same state with $f_{k,b}$ in place of $f'_{k,b}$.  Then
\begin{equation}
 \langle\phi|\phi'\rangle=1-\E_{b,x}H^2\bigl(f_{k,b}(x),f'_{k,b}(x)\bigr)\geq1-\eta_H,
 \qquad\text{so}\qquad
 \DeltaTr{\phi}{\phi'}\leq\sqrt{2\eta_H}.
\end{equation}  For $|\phi\rangle$, the argument of
Proposition~\ref{prop:exact-preparation} applies verbatim: the supports of $f_{k,b}(x)$ are disjoint for fixed
$b$, and $f_{k,0}(x_0)=f_{k,1}(x_1)$ for every claw, so measuring $Y$ leaves exactly the claw state of the
observed image.  For $|\phi'\rangle$, every outcome $(b,x,y)$ of nonzero probability satisfies $\Chk(k,b,x,y)=1$,
so its image is well formed by Lemma~\ref{lem:well-formed}.  Consequently, the honest preimage test still accepts
with probability one, the branch bits still agree exactly, and the equation test loses at most an additive
$\sqrt{2\eta_H}$ (Proposition~\ref{prop:completeness}).
\end{remark}

\subsection{Observables for an arbitrary device}\label{sec:explicit-meas}

Soundness does not assume the state or the honest measurements above.  After the joint device commits to $y$, let
$\{P_y^{b,x,\vec b}\}$ and $\{M_y^{r}\}_{r=(d,\vec u)}$ be its complete preimage- and equation-response
measurements, where $(b,x)$ is Alice's preimage answer and $\vec b=(b_2,\ldots,b_N)$ are the Bobs' branch bits. We may regard them as global projective measurements after adjoining work registers.
Alice's answer alone is described by the projectors
\begin{equation}\label{eq:alice-marginal}
 \Pi_y^{b,x}:=\sum_{\vec b}P_y^{b,x,\vec b},
\end{equation}
which leave the Bobs' preimage answers unmeasured.  Recall from Definition~\ref{def:ntcf}, item~3(b), that
$\Chk(k,b,x,y)=1$ exactly when $y\in\Supp(f'_{k,b}(x))$, and that $\Chk$ needs no trapdoor; and recall from
Lemma~\ref{lem:well-formed} that $x_b(y)=\Inv(\td,b,y)$ and that an image $y$ is well formed if
$\Chk(k,b,x_b(y),y)=1$ for some $b$.  By that lemma, Alice's answer is valid only if $y$ is well formed and
$x=x_b(y)$.  Hence the valid-preimage projectors are
\begin{align}
 \PiN{b}(y)&:=\sum_{x:\Chk(k,b,x,y)=1}\Pi_y^{b,x}=
 \begin{cases}\Pi_y^{b,x_b(y)}&\text{if $y$ is well formed},\\ 0&\text{otherwise},\end{cases}
 \qquad b\in\{0,1\},\label{eq:preimage-projectors}\\
 \PiN{\perp}(y)&:=\id-\PiN{0}(y)-\PiN{1}(y).\nonumber
\end{align}
For the equation response, recall from~\eqref{eq:good-set} that a well-formed image $y$ has the claw
$(x_0,x_1)=(x_0(y),x_1(y))\in\Rk$ and the set of admissible equation strings
$G_{k,y}=G_{k,0,x_0}\cap G_{k,1,x_1}\subseteq\{0,1\}^w$.  A uniformly random $d$ lies outside $G_{k,y}$ with
probability at most $2\eta_G(\lambda)$, and membership in $G_{k,y}$ can be decided with the trapdoor
(Definition~\ref{def:ntcf}, item~4).  Call an answer $r$ \emph{malformed} if it is not of the form
$(d,\vec u)\in\{0,1\}^w\times\{0,1\}^N$.  We partition the answers $r=(d,\vec u)$ of a well-formed image
into the classes
\begin{align}
 \sC_y&=\{r:\ d\in G_{k,y}\text{ and }{\textstyle\bigoplus_{j=1}^N}u_j=d\cdot(x_0\oplus x_1)\},\nonumber\\
 \sI_y&=\{r:\ d\in G_{k,y}\text{ and }{\textstyle\bigoplus_{j=1}^N}u_j\neq d\cdot(x_0\oplus x_1)\},
 \label{eq:equation-classes}\\
 \sB_y&=\{r:\ d\notin G_{k,y}\text{ or $r$ is malformed}\},\nonumber
\end{align}
and put $\sC_y=\sI_y=\emptyset$ if $y$ is not well formed.  Write $M_{\sC}(y)=\sum_{r\in\sC_y}M_y^r$, and similarly
$M_{\sI}(y)$ and $M_{\sB}(y)$.  The operational equation test accepts exactly the class $\sC_y$:
\begin{equation}\label{eq:equation-projector}
 \MN{0}(y):=M_{\sC}(y),\qquad \MN{1}(y):=\id-\MN{0}(y)=M_{\sI}(y)+M_{\sB}(y).
\end{equation}
The admissibility check is essential, since $d=0^w$ would otherwise make the equation trivial.  Alice's valid
branch is the privacy observable.  Agreement between Alice and Bob $j$ is a correctness property, described by the
disagreement projector
\begin{equation}\label{eq:disagreement-projector}
 D_{j,y}:=\sum_{b,x,\vec b:\ b_j\neq b}P_y^{b,x,\vec b};
\end{equation}
these projectors need not commute with the equation measurement and are never used in the two-measurement
entropy argument.

Without channel noise, the honest implementation fails the equation test only with negligible probability.
Proposition~\ref{prop:completeness} below bounds this probability by
\begin{equation}\label{eq:nuG-definition}
 \nuG(\lambda):=2\eta_G(\lambda)+\sqrt{2\eta_H(\lambda)}.
\end{equation}
The first term accounts for an inadmissible string $d$, with $\eta_G$ as in Definition~\ref{def:ntcf}, item~4.
The second accounts for the approximation error of the efficient range superposition, with $\eta_H$ the Hellinger
bound of item~3(c); it vanishes for noiseless families.

\begin{proposition}[Honest completeness of the two tests]\label{prop:completeness}
	In the honest implementation without channel noise, the preimage test accepts with probability one, the
	computational-basis branch bits of all parties agree with probability one, and the equation test accepts with
	probability at least $1-\nuG(\lambda)$.
\end{proposition}

\begin{proof}
	First consider a noiseless family.  Each computational-basis term of~\eqref{eq:target-state} contains a
	valid Alice preimage and identical branch bits, which proves the first two claims.  By
	Proposition~\ref{prop:xor}, the parity check always passes and $d$ is uniform, so the only rejection is the
	event $d\notin G_{k,y}$, which has probability at most $2\eta_G(\lambda)$ by~\eqref{eq:good-set}.  For a general
	family, Remark~\ref{rem:ntcf-honest} shows that every image is well formed and every preimage answer valid,
	and the branch bits are exact copies.  For the equation test, write $x_b(y)=\Inv(\td,b,y)$ and let $F$ be the
	event that $d\notin G_{k,0,x_0(y)}\cap G_{k,1,x_1(y)}$ or that the parity check fails for the pair
	$(x_0(y),x_1(y))$.  This event is defined for every image, well formed or not.  The honest test rejects exactly
	on $F$ or when $y$ is not well formed, and the latter has probability zero on $|\phi'\rangle$; hence the
	failure probability equals $\Pr_{\phi'}[F]$.  On the state $|\phi\rangle$ of Remark~\ref{rem:ntcf-honest}, every
	image of nonzero probability lies in $\Supp(f_{k,0}(x_0))=\Supp(f_{k,1}(x_1))$ for a claw $(x_0,x_1)$, so
	$x_b(y)=x_b$ by item~2 of Definition~\ref{def:ntcf}, and measuring $Y$ leaves the exact claw state.  As in the
	noiseless case, $\Pr_\phi[F]\leq2\eta_G$.  Since $F$ results from the same operations and measurements applied
	to either state,
	\[
	 \Pr_{\phi'}[F]\leq\Pr_\phi[F]+\DeltaTr{\phi}{\phi'}\leq2\eta_G+\sqrt{2\eta_H}=\nuG.
	\]
\end{proof}

\section{The DI-CKA Protocol}\label{sec:protocol}

\subsection{Model and trust boundary}\label{sec:model}\label{sec:trust-boundary}

\paragraph{Honest parties.}
Alice and the Bobs follow the protocol.  A trusted classical verifier $V$, hosted by Alice, holds private
randomness that is independent of the device.  Authenticated public classical channels connect Alice to each Bob;
no Bob-to-Bob channel is required.  

\paragraph{The device.}
The parties interact with an untrusted apparatus with nominal interfaces $D_1,\ldots,D_N$.  All interfaces, their
shared memory, and all classical or quantum communication between them, including the quantum links over which
branch registers travel in the honest implementation, form one joint machine $D$.  It may retain quantum memory
across rounds and adapt to the entire past, and it must be QPT during the challenge--response phase.  There is no
tensor-product, non-signaling, or space-like-separation assumption.  Tampering with the quantum links is part of
the device strategy.

\paragraph{The eavesdropper.}
Eve is computationally unbounded.  She may prepare the joint state $\rho_{DE}$ before the protocol, and she
observes the entire public transcript.  The device has no channel to Eve during the execution; in particular, the
communication between its components is inaccessible to her.  As emphasized in~\cite{Metger2024}, this is
unavoidable once the components may communicate: otherwise a malicious device could simply forward the key to
Eve, and no test could detect it.  After privacy amplification the device is isolated or destroyed.  Trapdoors,
images, preimages, and Alice's own equation answers are local data of the verifier and are erased after use.

\paragraph{Honest implementation.}
The honest device needs no communication between its components: by Remark~\ref{rem:ghz-preshared}, it can be
run with pre-distributed GHZ states and local operations.  In the language of~\cite{Metger2024}, the additional
communication permitted by our model is used only to describe attacks.  This excludes trivial protocols in which
an honest device would rely on a private channel between the components, and it allows our protocol to be
layered on top of a standard DI setting in which isolation of the components is attempted but not relied upon.

\subsection{Protocol}\label{sec:protocol-description}

\begin{protocolbox}[title={\textbf{Protocol 1: Computational DI Conference Key Agreement}}]
	\refstepcounter{protocol}\label{prot:dicka}%

	\textbf{Parameters.}
	Fix $n=\poly(\lambda)$ rounds, test probability $\gamma\in(0,1)$, conditional equation-test probability
	$\beta\in(0,1)$, target conditional test success $\omega_{\rm exp}$, estimation margin
	$\delta_{\rm est}>0$, test-count margin $\delta_{\rm samp}>0$, a minimum preimage-test sample size
	$m_p^{\min}$, Alice--Bob disagreement thresholds $Q_2^{\rm tol},\ldots,Q_N^{\rm tol}$, a reconciliation
	leakage bound $L_{\rm EC}$, and security parameters
	$\varepsilon_s,\varepsilon_{\rm PA},\varepsilon_{\rm corr},\varepsilon_{\rm acc}$.

	\medskip
	\textbf{Challenge--response phase.}
	All failures are recorded as private flags; no abort decision is made until all $n$ rounds are complete.  For
	each $i\in[n]$:
	\begin{enumerate}
		\item \emph{Commitment.} Alice samples $(k_i,\td_i)\leftarrow\Gen(1^\lambda)$, sends $k_i$ to the device,
		and receives an image $y_i$.  She computes $x_b^{(i)}:=\Inv(\td_i,b,y_i)$ for $b\in\{0,1\}$ and flags $y_i$
		as malformed unless it is well formed (Lemma~\ref{lem:well-formed}).

		\item \emph{Hidden round choice.} Alice privately samples
		$G_i\leftarrow\operatorname{Bernoulli}(1-\gamma)$, where $G_i=1$ denotes generation.  She sets $T_i=0$ if
		$G_i=1$ and otherwise samples $T_i\leftarrow\operatorname{Bernoulli}(\beta)$.  She announces only $T_i$,
		so generation and preimage-test rounds present the same challenge $T_i=0$.

		\item \emph{Response and validation.} If $T_i=0$, the device returns $(b_1^{(i)},x^{(i)})$ to Alice and
		$b_j^{(i)}$ to each Bob.  Alice sets
		\begin{equation}\label{eq:valid-preimage-flag}
			V_i=1\quad\Longleftrightarrow\quad
			y_i\text{ is well formed and }
			\Chk(k_i,b_1^{(i)},x^{(i)},y_i)=1.
		\end{equation}
		This check is performed in every round with $T_i=0$, including generation rounds: since $G_i$ is still
		hidden, generation and preimage-test rounds are treated identically up to this point.
		If $T_i=1$, the device returns $(d^{(i)},u_1^{(i)})$ to Alice and $u_j^{(i)}$ to each Bob, who forwards it
		to Alice over the public channel.  Alice sets $V_i=1$ if and only if $y_i$ is well formed and
		\begin{equation}\label{eq:protocol-equation-check}
			d^{(i)}\in G_{k_i,y_i}
			\quad\text{and}\quad
			\bigoplus_{j=1}^{N}u_j^{(i)}
			=d^{(i)}\cdot(x_0^{(i)}\oplus x_1^{(i)}).
		\end{equation}
		Otherwise $V_i=0$.

		\item \emph{Delayed disclosure.} After the response has been recorded, Alice announces $G_i$.  If $G_i=0$,
		she announces the test result $W_i:=V_i$; if moreover $T_i=0$, Alice announces $b_1^{(i)}$, each Bob
		announces $b_j^{(i)}$, and the parties record $D_{j,i}:=b_1^{(i)}\oplus b_j^{(i)}$.  If $G_i=1$, the parties
		keep their branch bits as raw-key symbols and set $W_i=\perp$.  Alice keeps $V_i$ private; it records whether
		the generation preimage is valid and is used only in check~2 of the parameter estimation below.

		\item \emph{Erasure.} Alice erases $\td_i$, $y_i$, $x^{(i)}$, and, if $T_i=1$, her answer
		$(d^{(i)},u_1^{(i)})$, once $V_i$ has been recorded.
	\end{enumerate}

	\textbf{Deferred parameter estimation.}
	Define the preimage-test sample
	\begin{equation}
		\mathcal I_p:=\{i:G_i=0,\ T_i=0\},\qquad m_p:=|\mathcal I_p|.
	\end{equation}
	After all $n$ rounds, abort unless:
	\begin{enumerate}
		\item the total test score and the number of test rounds satisfy
		\begin{align}
			\sum_{i:G_i=0}W_i&\ge(\gamma\omega_{\rm exp}-\delta_{\rm est})n,
			\label{eq:win-count-condition}\\
			|\{i:G_i=0\}|&\le(\gamma+\delta_{\rm samp})n;
			\label{eq:test-count-cap}
		\end{align}
		\item every generation round has $V_i=1$; and
		\item $m_p\ge m_p^{\min}$ and, for every $j=2,\ldots,N$,
		\begin{equation}\label{eq:disagreement-test}
			\sum_{i\in\mathcal I_p}D_{j,i}\le Q_j^{\rm tol}m_p.
		\end{equation}
	\end{enumerate}

	\medskip
	\textbf{Reconciliation, verification, and privacy amplification.}
	Let $Z_A=(b_1^{(i)}:G_i=1)$ and $Z_j=(b_j^{(i)}:G_i=1)$.  Alice broadcasts a one-way reconciliation message,
	from which Bob $j$ computes an estimate $\widetilde Z_j$ of $Z_A$.  The combined syndrome-and-verification
	transcript has support of size at most $2^{L_{\rm EC}}$.  The protocol aborts on a failed verification; the tags
	are chosen so that, with $\Omega$ the event that the protocol does not abort (Section~\ref{sec:security-def}),
	\begin{equation}\label{eq:verification-correctness}
		\Pr[\Omega\wedge\exists j\ge2:\widetilde Z_j\ne Z_A]
		\le\varepsilon_{\rm corr}.
	\end{equation}
	Finally, Alice publishes a seed for a two-universal hash function, and all parties hash their reconciled strings
	to $\ell$-bit keys: Alice obtains $K_A$ from $Z_A$, and Bob $j$ obtains $K_j$ from $\widetilde Z_j$.

	\medskip
	\textbf{Public transcript.}
	The transcript $C$ consists of the keys $k_i$, the challenges $T_i$, the flags $G_i$, the announcements of
	step~4, the Bobs' forwarded equation answers, the abort decision, and the reconciliation, verification, and
	hashing messages.  Images, preimages, trapdoors, and Alice's equation answers never become public.
\end{protocolbox}

\begin{remark}[Necessity of timing and checks]\label{rem:protocol-timing}
	The delayed announcement of $G_i$ forces generation and preimage-test rounds to use the same physical response
	instrument.  If $G_i$ were announced before a $T_i=0$ response, the device could use one measurement in tested
	preimage rounds and an unrelated, fully predictable measurement in generation rounds.  The AHB relation would
	then constrain only the former and certify no entropy in the latter.

	Likewise, accepting an arbitrary $d$ in~\eqref{eq:protocol-equation-check} would make the equation test
	vacuous, and aborting immediately on a malformed image or invalid generation preimage would reveal a variable
	stopping time to the device.  The admissibility check and the deferred abort rule close these two loopholes.
\end{remark}

\begin{proposition}[Noiseless honest completeness]\label{prop:protocol-completeness}
Let $\Omega$ denote the event that Protocol~\ref{prot:dicka} does not abort.  Suppose the honest implementation
is independent across rounds and has no channel noise.  Suppose that $m_p^{\min}\le(\gamma(1-\beta)-\delta_p)n$ for
some $\delta_p>0$, and that
\begin{equation}
 \Delta_{\rm win}:=\gamma(1-\beta\nuG(\lambda)-\omega_{\rm exp})+\delta_{\rm est}>0.
\end{equation}
Then
\begin{equation}\label{eq:honest-completeness-bound}
 \Pr[\neg\Omega]\leq e^{-2n\Delta_{\rm win}^2}
 +e^{-2n\delta_{\rm samp}^2}+e^{-2n\delta_p^2}+\varepsilon_{\rm post},
\end{equation}
where $\varepsilon_{\rm post}$ bounds the failure probability of honest post-processing.
\end{proposition}

\begin{proof}
	Set $X_i:=\mathbf 1\{G_i=0,\ W_i=1\}$, $Y_i:=\mathbf 1\{G_i=0\}$, and
	$Z_i:=\mathbf 1\{G_i=0,T_i=0\}$.  The triples $(X_i,Y_i,Z_i)$ are independent across rounds.  By
	Proposition~\ref{prop:completeness}, an honest preimage test always succeeds, whereas an honest equation test
	fails with probability at most $\nuG(\lambda)$.  Consequently,
	\[
	\mathbb E X_i\ge p_{\rm win}:=\gamma(1-\beta\nuG(\lambda)),\qquad
	\mathbb E Y_i=\gamma,\qquad
	\mathbb E Z_i=p_p:=\gamma(1-\beta).
	\]
	Let $\mathcal A_{\rm win}$, $\mathcal A_{\rm count}$, and $\mathcal A_p$ denote failure of, respectively, the
	win-count, test-count, and minimum-sample checks.  The corresponding thresholds differ from the above means by
	at least $\Delta_{\rm win}$, $\delta_{\rm samp}$, and $\delta_p$, so Hoeffding's inequality gives
	\[
	\Pr[\mathcal A_{\rm win}]\le e^{-2n\Delta_{\rm win}^2},\qquad
	\Pr[\mathcal A_{\rm count}]\le e^{-2n\delta_{\rm samp}^2},\qquad
	\Pr[\mathcal A_p]\le e^{-2n\delta_p^2}.
	\]
	In the noiseless honest implementation every generation preimage is valid and every $D_{j,i}$ vanishes, so no
	other parameter-estimation check fails.  A union bound, together with the post-processing failure probability
	$\varepsilon_{\rm post}$, proves~\eqref{eq:honest-completeness-bound}.
\end{proof}

\subsection{Main theorem}\label{sec:main-guarantee}

\begin{theorem}[Computational DI-CKA security]\label{thm:main}
Consider Protocol~\ref{prot:dicka} in the model of Section~\ref{sec:model}, instantiated with an NTCF family
(Definition~\ref{def:ntcf}) that has the adaptive hardcore bit property against QPT adversaries holding
polynomial-size quantum advice (Definition~\ref{def:ahb-2party}), as guaranteed by Assumption~\ref{ass:ntcf}.

Let $g_\beta(\omega)$ be the certified single-round entropy: a lower bound, valid for every efficient device up to a
negligible term $\xi_{\rm AHB}(\lambda)$, on the entropy of Alice's preimage outcome given Eve's information, when a
test round passes with probability $\omega$.  Let $a_0\geq0$ and $b_0$ be such that
$a_0\omega+b_0\leq g_\beta(\omega)$ for all $\omega\in[0,1]$.  Assume $\delta_{\rm est}\leq\gamma\omega_{\rm exp}$, fix
$\varepsilon_s,\varepsilon_{\rm acc},\varepsilon_{\rm PA}\in(0,1)$, and set
\begin{align}
 t&:=(1-\beta\gamma)\left[a_0\left(\omega_{\rm exp}-\frac{\delta_{\rm est}}{\gamma}\right)+b_0-\xi_{\rm AHB}(\lambda)\right],
 \label{eq:thm-t}\\
 v&:=2\left(\log13+\left\lceil\frac{(1-\beta\gamma)a_0}{\gamma}\right\rceil\right)
 \sqrt{1-2\log(\varepsilon_s\varepsilon_{\rm acc})},\label{eq:thm-v}\\
 L_{\rm drop}&:=(\gamma+\delta_{\rm samp})\,n\log3,\qquad
 L_{\rm test}:=(\gamma+\delta_{\rm samp})\,nN.\label{eq:thm-leak}
\end{align}
If the key length satisfies
\begin{equation}\label{eq:key-length}
 \ell\leq nt-v\sqrt n-L_{\rm drop}-L_{\rm test}-L_{\rm EC}
 -2\log\frac1{2\varepsilon_{\rm PA}},
\end{equation}
where $L_{\rm EC}$ is the reconciliation leakage bound of Protocol~\ref{prot:dicka}, then the protocol is
$\varepsilon_{\rm corr}$-correct and $(\varepsilon_{\rm acc}+2\varepsilon_s+\varepsilon_{\rm PA})$-secret, so its total
error is at most $\varepsilon_{\rm corr}+\varepsilon_{\rm acc}+2\varepsilon_s+\varepsilon_{\rm PA}$.
\end{theorem}

In~\eqref{eq:key-length}, $nt$ is the entropy certified by the tests, $v\sqrt n$ is the finite-size correction of
entropy accumulation, $L_{\rm drop}$ and $L_{\rm test}$ account for the publicly announced test data, $L_{\rm EC}$
for reconciliation and verification, and the last term for privacy amplification.  The quantities in the theorem
are derived in the proof as follows.  The function $g_\beta$ is defined in~\eqref{eq:g-beta} from the single-round
bound $g$ of~\eqref{eq:mp-entropy} and Theorem~\ref{thm:single-round}, by minimizing over all splits between the two
tests that are compatible with $\omega$.  The affine lower bound $a_0\omega+b_0$ (the function
$h_{\omega_0}$ of Section~\ref{sec:main}) yields the min-tradeoff function~\eqref{eq:fmin-def} (Lemma~\ref{lem:min-tradeoff}), and $t$ is its minimum over the acceptance
set~\eqref{eq:t-definition}, evaluated in Appendix~\ref{app:eat-details}.  The constant $v$ is the second-order term
of the EAT~\cite{DFR2020} in the form~\eqref{eq:eat-v}, with output alphabet size $d_O\leq6$
(Lemma~\ref{lem:appendix-eat-channels}) and gradient $(1-\beta\gamma)a_0/\gamma$.  Finally, $L_{\rm drop}$ and
$L_{\rm test}$ are the bounds~\eqref{eq:Ldrop} and~\eqref{eq:explicit-Ltest}.

The rate $t$ contains the negligible term $\xi_{\rm AHB}(\lambda)$, as does every finite-size statement based on an
asymptotic hardness assumption.  A concrete-security instantiation would bound it through the LWE parameters;
cf.\ the discussion in~\cite{MA2025}.

\begin{corollary}[Two-party specialization]\label{cor:diqkd}
For $N=2$, Protocol~\ref{prot:dicka} is a computational DI-QKD protocol with the parameters and security error
of Theorem~\ref{thm:main}.  Its proof permits arbitrary communication between the two nominal device interfaces
and arbitrary non-IID memory across rounds.
\end{corollary}

\begin{proof}
With one Bob, conference key agreement is two-party key agreement.  The collective parity becomes
$u_1\oplus u_2$, and every reduction and entropy bound below remains valid without modification.  The carried
register in the EAT contains both interfaces and their complete memory, so the specialization retains security
against coherent attacks and makes no non-signaling assumption.
\end{proof}

The proof of Theorem~\ref{thm:main} is modular.  Section~\ref{sec:mp-ahb} reduces the collective parity premise
to ordinary AHB, Section~\ref{sec:mp-goodsub} converts it into complementary binary measurements,
Section~\ref{sec:single-round} derives the one-round entropy bound, and Section~\ref{sec:main} accumulates this
entropy and charges every public leakage term.  Section~\ref{sec:lifting} then establishes the post-execution
everlasting guarantee.

\section{Collective Adaptive Hardcore Bit}\label{sec:mp-ahb}

This section isolates the only place where the computational assumption enters: we define the collective AHB
property for a jointly operating device and show that it follows from the ordinary AHB property without loss.

Fix a fresh public key and run the joint QPT device until it reports an image $y$.  Measure Alice's preimage
answer with $\{\Pi_y^{b,x}\}_{b,x}$ from~\eqref{eq:alice-marginal}, leaving the Bobs' preimage answers unmeasured,
and then apply the equation-response measurement $\{M_y^r\}$ to the residual state.  This sequential experiment
is analytical; the protocol asks only one of the two questions in each round.  It is efficiently implementable:
one applies the dilated preimage instrument coherently, measures only Alice's output registers, and uncomputes
the dilation (Appendix~\ref{app:collective-ahb}).  

\begin{definition}[Collective AHB property]\label{def:mp-ahb}
Recall from~\eqref{eq:equation-classes} that, for a well-formed image $y$ with claw $(x_0,x_1)$, an equation
answer $r=(d,\vec u)$ lies in $\sC_y$ if $d\in G_{k,y}$ and $\bigoplus_{j=1}^Nu_j=d\cdot(x_0\oplus x_1)$, in $\sI_y$
if $d\in G_{k,y}$ and the parity is incorrect, and in $\sB_y$ otherwise.  In the sequential experiment, let $\sC$
be the event that $y$ is well formed, Alice's answer $(b,x)$ satisfies $\Chk(k,b,x,y)=1$, and the equation answer
lies in $\sC_y$; define $\sI$ in the same way with $\sI_y$ in place of $\sC_y$.  The collective advantage of a joint device $\cD$ is
\begin{equation}\label{eq:collective-advantage}
 \AdvN(\cD):=\bigl|\Pr[\sC]-\Pr[\sI]\bigr|,
\end{equation}
where the probabilities include the generation of the fresh key.  Invalid preimages, malformed images, and answers
in $\sB_y$ belong to neither event.  The family $\cF$ has the $N$-party collective AHB property if $\AdvN(\cD)$ is
negligible for every joint QPT device $\cD$.
\end{definition}

This is the direct multiparty generalization of the AHB property: the single predicted hardcore bit is replaced by
the parity of the $N$ equation answers.  No condition $b_j=b$ is imposed on the Bobs' preimage answers; that
condition belongs to correctness, while the ordinary AHB game contains only Alice's valid preimage.

\begin{theorem}[Collective AHB reduces to ordinary AHB]\label{thm:mp-ahb}
For every joint QPT device $\cD$ for $N$ parties there is a QPT adversary $\cA^*$ for the ordinary (two-party) AHB
game of Definition~\ref{def:ahb-2party}, which runs $\cD$ as a subroutine and uses the same advice, such that
\begin{equation}\label{eq:exact-reduction}
 \bigl|\Pr[\cA^*\in H_k]-\Pr[\cA^*\in\bar H_k]\bigr|=\AdvN(\cD).
\end{equation}
In particular, if the family has the AHB property (Definition~\ref{def:ahb-2party}), then it has the $N$-party
collective AHB property, with $\AdvN(\cD)\leq\eta_{\rm AHB}(\lambda)$ for the negligible function $\eta_{\rm AHB}$
associated with $\cA^*$.  The reduction is exact and incurs no loss depending on $N$.
\end{theorem}

\begin{proof}
We first isolate the transcript-level fact used by the reduction.

\begin{lemma}[Event preservation]\label{lem:event-preservation}
For a well-formed image, a valid Alice preimage, and an equation answer outside $\sB_y$, the map
\begin{equation}\label{eq:transcript-map}
 (y,b,x,d,u_1,\ldots,u_N)\longmapsto(b,x,d,e),
 \qquad e=\bigoplus_{j=1}^Nu_j,
\end{equation}
maps membership in $\sC$ and $\sI$ exactly to membership in $H_k$ and $\bar H_k$, respectively.
\end{lemma}

\begin{proof}
By Lemma~\ref{lem:well-formed}, a valid answer satisfies $x=x_b(y)$, and $(x_0(y),x_1(y))\in\Rk$ is the claw
containing it, so $G_{k,y}=G_{k,0,x_0}\cap G_{k,1,x_1}$ is exactly the admissible set in $H_k$.  The map preserves
$b$, $x$, and $d$, and its final bit satisfies $e=d\cdot(x_0\oplus x_1)$ exactly when the collective parity is
correct.  Negating this equality proves the incorrect case.
\end{proof}

Construct an ordinary AHB adversary $\cA^*$.  Given the fresh key and key-independent advice, it runs the
complete joint strategy, including every nominal interface and all internal communication, and performs the
sequential experiment above.  If the image is well formed (which $\cA^*$ cannot check, but which every valid
answer certifies by Lemma~\ref{lem:well-formed}), Alice's answer is valid, and the equation answer is well
typed, it outputs the tuple~\eqref{eq:transcript-map}.  On any other outcome it outputs some well-typed prefix
$(b^\star,x^\star,d^\star)$ together with a prediction bit $e$ that is uniformly random and independent of everything
else.  Since $\Chk$ requires no trapdoor, $\cA^*$ can decide the validity of Alice's answer.

Consider the three cases.  If the original answer is valid and lies outside $\sB_y$,
Lemma~\ref{lem:event-preservation} preserves the two events.  If it is valid but lies in $\sB_y$ because
$d\notin G_{k,y}$, neither experiment counts it.  Finally, fix a fallback prefix $(b^\star,x^\star,d^\star)$.  If it is
not in the support of $H_k\cup\bar H_k$, neither value of $e$ is counted; otherwise exactly one value of $e$ is
correct and one is incorrect.  Thus, if $p_{\rm fb}$ is the probability of an eligible fallback prefix,
\begin{align}
 \Pr[\cA^*\in H_k]&=\Pr[\sC]+p_{\rm fb}/2,\\
 \Pr[\cA^*\in\bar H_k]&=\Pr[\sI]+p_{\rm fb}/2.
\end{align}
The fallback terms cancel upon subtraction, which proves~\eqref{eq:exact-reduction}.  Since $N=\poly(\lambda)$, the
reduction remains QPT, and the second claim follows from~\eqref{eq:ordinary-ahb}.
\end{proof}

The event probabilities are traces of sequential post-measurement states.  After commitment, write the state as
$\Phi_{KYDE}=\sum_{k,y}p(k,y)|k,y\rangle\!\langle k,y|\otimes\Phi^{k,y}_{DE}$.  By~\eqref{eq:preimage-projectors},
a valid answer $(b,x)$ corresponds to the projector $\PiN{b}(y)$, so
\begin{equation}\label{eq:ahb-trace-form}
 \Pr[\sC]=\sum_{k,y}p(k,y)\sum_{b}\Tr\!\left(M_{\sC}(y)\PiN{b}(y)\Phi^{k,y}\PiN{b}(y)\right),
\end{equation}
and analogously for $\Pr[\sI]$ with $M_{\sI}(y)$.  This is the form in which collective AHB enters the operator
argument.  The advantage is averaged over the fresh key and the image, and we never need a bound for a fixed pair
$(k,y)$.

\paragraph{Uniformity over histories.}
For round $i$, the incoming device state may depend on the entire preceding transcript and may be inefficient to
prepare.  It is nevertheless a polynomial-size state fixed before the fresh key $k_i$ is sampled, and is
therefore valid key-independent advice in Definition~\ref{def:ahb-2party}; the round index can be included in the
advice as well.  Since $\eta_{\rm AHB}$ does not depend on the advice, Theorem~\ref{thm:mp-ahb} holds uniformly over
every history encountered by entropy accumulation.

\section{Binary Measurements and the Good Subspace}\label{sec:mp-goodsub}

This section turns the collective AHB property into the operator statement used for entropy: after a balanced
completion of the equation test, almost all of the device's state lies in blocks where the two tests are
complementary.

The equation answers split into the classes $\sC_y$, $\sI_y$, and $\sB_y$ of~\eqref{eq:equation-classes}.  The
AHB property controls $\sC_y$ against $\sI_y$, but the operational test rejects $\sI_y\cup\sB_y$.  Adjoin an
independent uniformly random coin in a register $\Coin$ and define the analytical binary measurement
\begin{align}
 \MBN{0}(y)&=M_{\sC}(y)\otimes\id_{\Coin}+M_{\sB}(y)\otimes|0\rangle\!\langle0|_{\Coin},
 \label{eq:balanced-equation-zero}\\
 \MBN{1}(y)&=M_{\sI}(y)\otimes\id_{\Coin}+M_{\sB}(y)\otimes|1\rangle\!\langle1|_{\Coin}.
 \label{eq:balanced-equation-one}
\end{align}
It is used only in the analysis; the protocol continues to reject every answer in $\sB_y$.

Recall from~\eqref{eq:preimage-projectors} that, for $b\in\{0,1\}$, $\PiN{b}(y)$ is the projector onto Alice
returning the valid preimage $(b,x_b(y))$ of the image $y$, and that it vanishes if $y$ is not well formed.  Recall
also from~\eqref{eq:equation-projector} that $\MN{0}(y)=M_{\sC}(y)$ is the projector onto passing the equation
test.  Let $\Phi_{KYDE}$ be the state after a fresh key and image, let $P_Y=\PiN{0}(Y)+\PiN{1}(Y)$ be the projector
onto a valid preimage answer, and let
\begin{equation}\label{eq:valid-projected-state}
\begin{aligned}
 \omega_p&=\Tr(P_Y\Phi),&\qquad
 \omega_m&=\Tr(\MN{0}(Y)\Phi),\\
 \bar\omega_m&=\Tr\bigl(\MBN{0}(Y)(\Phi\otimes\tau_{\Coin})\bigr),&\qquad
 \Psi&=\frac{P_Y\Phi P_Y}{\omega_p},
\end{aligned}
\end{equation}
where $\tau_{\Coin}$ is the maximally mixed coin state.  Thus $\omega_p$ and $\omega_m$ are the success
probabilities of the preimage and equation tests, and $\Psi$ is the state conditioned on a valid preimage.  We
suppress the tensor factor $\tau_{\Coin}$ when no confusion arises.

\begin{lemma}[Balanced completion]\label{lem:balanced-completion}
The sequential bias of the balanced measurement on $\Psi$,
\begin{equation}\label{eq:balanced-mu}
 \mu:=\left|\frac12-\Tr\!\left(\MBN{0}\PiN{0}\Psi\PiN{0}\right)-\Tr\!\left(\MBN{0}\PiN{1}\Psi\PiN{1}\right)\right|,
\end{equation}
satisfies
\begin{equation}\label{eq:mu-bound}
 \mu=\frac{|\Pr[\sC]-\Pr[\sI]|}{2\omega_p}\leq\frac{\eta_{\rm AHB}(\lambda)}{2\omega_p}.
\end{equation}
Moreover, $\bar\omega_m\geq\omega_m$.
\end{lemma}

\begin{proof}
Since $\PiN{b}P_Y=\PiN{b}$, we have $\PiN{b}\Psi\PiN{b}=\PiN{b}\Phi\PiN{b}/\omega_p$.  Tracing out the coin, the
balanced measurement assigns half of the class $\sB$ to each outcome, so by~\eqref{eq:ahb-trace-form}
\begin{align}
 \sum_b\Tr\!\left(\MBN{0}\PiN{b}\Psi\PiN{b}\right)&=\frac{\Pr[\sC]+\frac12p_{\sB}}{\omega_p},\\
 \sum_b\Tr\!\left(\MBN{1}\PiN{b}\Psi\PiN{b}\right)&=\frac{\Pr[\sI]+\frac12p_{\sB}}{\omega_p},
\end{align}
where $p_{\sB}$ is the probability that Alice's answer is valid and the subsequent equation answer lies in $\sB$.
The three classes partition all equation answers, so $\Pr[\sC]+\Pr[\sI]+p_{\sB}=\omega_p$, and the quantity inside
the absolute value in~\eqref{eq:balanced-mu} equals $(\Pr[\sI]-\Pr[\sC])/(2\omega_p)$.  Theorem~\ref{thm:mp-ahb}
proves~\eqref{eq:mu-bound}.  Finally,
$\bar\omega_m=\Tr((M_{\sC}+\frac12M_{\sB})\Phi)\geq\Tr(M_{\sC}\Phi)=\omega_m$.
\end{proof}

The balanced completion is what makes the computational input to the operator argument exactly the AHB advantage.
If all of $\sB$ were assigned to rejection,
the corresponding bias would equal $|\Pr[\sC]-\Pr[\sI]-p_{\sB}|/(2\omega_p)$.  Its term $p_{\sB}$, the weight of
inadmissible answers after a preimage measurement, is not controlled by the AHB property, which compares only
$\sC$ with $\sI$.

The bound~\eqref{eq:mu-bound} is needed only when $\omega_p$ is close to one.  The entropy function of
Section~\ref{sec:single-round} vanishes unless $1-\omega_p<20^{-4}$, and in that regime
$\mu\leq\eta_{\rm AHB}(\lambda)$.

On $\Supp(P_Y)\otimes\mathcal H_{\Coin}$, set
\begin{equation}\label{eq:goodsub-projectors}
 \Pi_{\rm pre}=\PiN{0}(Y)\otimes\id_{\Coin},
 \qquad M_{\rm eq}=\MBN{0}(Y).
\end{equation}
Their complements on the projected support are $\PiN{1}(Y)\otimes\id_{\Coin}$ and $\MBN{1}(Y)$.  Jordan's lemma
gives
\begin{equation}\label{eq:protocol-jordan-decomposition}
 \Supp(P_Y)\otimes\mathcal H_{\Coin}=\bigoplus_\alpha\mathcal J_\alpha,
 \qquad\dim\mathcal J_\alpha\leq2,
\end{equation}
with every block invariant under both global projections.  Classical projectors onto $(K,Y)$ commute with both
operators, so the decomposition may refine those registers.  Let $c_\alpha\in[1/2,1]$ be the block overlap and
$\Gamma_Y(c)$ the projector onto the blocks with $c_\alpha\leq c$.

\begin{lemma}[Good-subspace projection~\cite{BCMVV2018,MA2025}]\label{lem:goodsub}
Let $\Pi,M$ be projections, $\psi$ a normalized state, and
\begin{equation}\label{eq:goodsub-mu}
 \mu=\left|\frac12-\Tr(M\Pi\psi\Pi)-\Tr(M(\id-\Pi)\psi(\id-\Pi))\right|.
\end{equation}
If $\omega=\Tr(M\psi)$ and $\Gamma(c)$ projects onto the Jordan blocks of square overlap at most $c$, then
\begin{equation}\label{eq:goodsub-exact}
 \Tr((\id-\Gamma(c))\psi)\leq\frac{2\mu+10\sqrt{1-\omega}}{(2c-1)^2}.
\end{equation}
\end{lemma}

For the projections~\eqref{eq:goodsub-projectors}, the bias in~\eqref{eq:goodsub-mu} is exactly~\eqref{eq:balanced-mu},
which Lemma~\ref{lem:balanced-completion} bounds.  It remains to express $\omega=\Tr(M_{\rm eq}\Psi)$ through the
observed quantities.  Gentle projection and the subadditivity of the square root give
\begin{align}
 \sqrt{\Tr(\MBN{1}(Y)\Psi)}
 &\leq2(1-\omega_p)^{1/4}+\sqrt{1-\omega_m},\label{eq:projected-equation-failure}\\
 \Tr(\MBN{0}(Y)\Psi)&\geq\omega_m-2\sqrt{1-\omega_p}.
 \label{eq:projected-equation-success}
\end{align}
Here $\omega_m$ is the operational equation success; Lemma~\ref{lem:balanced-completion} permits replacing the
unobserved balanced success $\bar\omega_m$ by this smaller value.  With $A(c)=10/(2c-1)^2$, substitution
into~\eqref{eq:goodsub-exact} yields
\begin{equation}\label{eq:bad-weight}
 \Tr((\id-\Gamma_Y(c))\Psi)
 \leq q_c+A(c)\frac{\mu}{5},\qquad
 q_c=A(c)\bigl(2(1-\omega_p)^{1/4}+\sqrt{1-\omega_m}\bigr).
\end{equation}
Thus successful tests force almost all weight onto blocks on which the preimage and equation measurements are
sufficiently complementary.  Appendix~\ref{app:goodsub-proof} proves the abstract lemma, and
Appendix~\ref{app:single-round} supplies the conditioning details behind~\eqref{eq:projected-equation-failure}
and~\eqref{eq:projected-equation-success}.

\section{Single-Round Entropy}\label{sec:single-round}

This section converts the good-subspace bound into an explicit lower bound on the entropy of Alice's preimage
outcome in a single round, as a function of the two observed test success probabilities.

Fix a polynomial-size grid $\mathcal C_\lambda\subset(1/2,1]$ with $2c-1\geq1/\poly(\lambda)$.  For
$c\in\mathcal C_\lambda$ define
\begin{align}
 q_c&=\frac{10}{(2c-1)^2}\left(2(1-\omega_p)^{1/4}+\sqrt{1-\omega_m}\right),\label{eq:q-c}\\
 z_c&=\omega_m-2\sqrt{1-\omega_p}-q_c,\label{eq:z-c}\\
 \kappa_p&=\sqrt{1-\omega_p}\log3+
 (1+\sqrt{1-\omega_p})\hbin\!\left(\frac{\sqrt{1-\omega_p}}{1+\sqrt{1-\omega_p}}\right),
 \label{eq:kappa-p}
\end{align}
and set
\begin{equation}\label{eq:mp-entropy}
 g(\omega_p,\omega_m)=\max\!\left\{0,
 \max_{\substack{c\in\mathcal C_\lambda\\z_c\in[1/2,1]}}
 (1-q_c)_+\bigl(\log(1/c)-\hbin(z_c)\bigr)-\kappa_p\right\}.
\end{equation}
Here $(x)_+:=\max\{x,0\}$ denotes the positive part.  In the proof, $(1-q_c)_+$ lower-bounds the weight of the
state on Jordan blocks of overlap at most $c$, and conditioned on these blocks the entropy is at least
$\log(1/c)-\hbin(z_c)$.
The maximization is restricted in two ways, so that the error term in Theorem~\ref{thm:single-round} below is the
same negligible function for every device.  First, the overlap $c$ ranges over a fixed polynomial-size grid with
$2c-1\geq1/\poly(\lambda)$.  A union bound over the grid then keeps the error negligible, even after it is
multiplied by $A(c)=10/(2c-1)^2$ (Appendix~\ref{app:uniform-opt}).  Second, $z_c$ is restricted to $[1/2,1]$,
where the binary entropy is decreasing, so a lower bound on a success probability gives an upper bound on its
entropy.  The function $g$ is nondecreasing in $\omega_m$, so the balanced success may safely be replaced by the smaller operational
equation success.  Since $A(c)\geq10$, the factor $(1-q_c)_+$ vanishes unless $1-\omega_p<20^{-4}$.

\begin{theorem}[Single-round entropy]\label{thm:single-round}
For every efficient joint device,
\begin{equation}\label{eq:single-round-bound}
 H(\widehat\Pi|EYK)_\Phi\geq g(\omega_p,\omega_m)-\xi_{\rm AHB}(\lambda),
\end{equation}
where $\widehat\Pi\in\{0,1,\perp\}$ is Alice's coarse preimage outcome and $\xi_{\rm AHB}$ is negligible.  The
bound has no explicit dependence on $N$.
\end{theorem}

\begin{proof}
If $g(\omega_p,\omega_m)=0$ there is nothing to prove, since the conditional entropy of a classical register is
nonnegative; hence assume $g>0$, so that $1-\omega_p<20^{-4}$.  We give the five steps that determine every term
in~\eqref{eq:mp-entropy}.  The valid-preimage and balanced equation projectors are controlled by the classical
registers $K,Y$:
\begin{equation}
 P_Y=\PiN{0}(Y)+\PiN{1}(Y),\qquad M_Y=\MBN{0}(Y).
\end{equation}
Keeping $K,Y$ inside the state is important: AHB bounds the experiment averaged over the fresh key and image, not
necessarily every fixed pair $(k,y)$.

\emph{Valid-preimage projection.}
Coarse-graining the complete preimage response to the three projectors $\PiN{0}$, $\PiN{1}$, and $\PiN{\perp}$
preserves the outcome $\widehat\Pi$ and its entropy.  On the valid support, the first two projectors form a binary measurement.  By
Lemma~\ref{lem:gentle}, the state $\Psi=P_Y\Phi P_Y/\omega_p$ is within trace distance
\begin{equation}\label{eq:epsilon-p}
 \epsilon_p=\sqrt{1-\omega_p}
\end{equation}
of $\Phi$.  For the equation response we use the balanced measurement, whose success is at least $\omega_m$.
The estimate $|\Tr A(\rho-\sigma)|\leq\|A\|_\infty\|\rho-\sigma\|_1$ gives
\begin{align}
 \Tr((\id-M_Y)\Psi)&\leq1-\omega_m+2\sqrt{1-\omega_p},\label{eq:main-projected-failure}\\
 \Tr(M_Y\Psi)&\geq\omega_m-2\sqrt{1-\omega_p}.\label{eq:main-projected-success}
\end{align}
Taking the square root of the first inequality and using $\sqrt{r+s}\leq\sqrt r+\sqrt s$ yields
\begin{equation}\label{eq:main-fourth-root}
 \sqrt{\Tr((\id-M_Y)\Psi)}
 \leq\sqrt{1-\omega_m}+2(1-\omega_p)^{1/4}.
\end{equation}
This second projection inside the good-subspace estimate is the source of the fourth-root loss.

\emph{Bad Jordan blocks.}
Apply Jordan's lemma to $\Pi_{\rm pre}$ and $M_{\rm eq}$.  Lemma~\ref{lem:balanced-completion} bounds their
sequential bias by $\mu\leq\eta_{\rm AHB}(\lambda)$, because $\omega_p\geq1/2$.  Equation~\eqref{eq:main-fourth-root}
and the good-subspace lemma then give~\eqref{eq:bad-weight}.  Thus, up to the negligible term $A(c)\mu/5$, the state
lies in blocks of overlap at most $c$ with probability at least $(1-q_c)_+$.

\emph{Equation success on good blocks.}
For events $A,B$, the inequality $\Pr[A\cap B]\geq\Pr[A]+\Pr[B]-1$ and~\eqref{eq:main-projected-success} give,
after selecting the good Jordan blocks,
\begin{equation}
 \Pr[M_{\rm eq}=0\mid\Gamma_Y(c)=1]\geq
 \omega_m-2\sqrt{1-\omega_p}-q_c-\negl(\lambda)=z_c-\negl(\lambda).
\end{equation}
We admit only $z_c\in[1/2,1]$, where the binary entropy decreases with the success probability.

\emph{Uncertainty.}
On every selected block the square overlap is at most $c$.  The uncertainty relation~\eqref{eq:eur} and the
binary entropy of the equation outcome imply
\begin{equation}
 H(\widehat\Pi|EYK,\Gamma=1)_\Psi\geq\log(1/c)-\hbin(z_c)-\negl(\lambda).
\end{equation}
Conditioning on the classical block flag and discarding the nonnegative contribution of the bad blocks yields
\begin{equation}\label{eq:projected-entropy}
 H(\widehat\Pi|EYK)_\Psi
 \geq(1-q_c)_+\bigl(\log(1/c)-\hbin(z_c)\bigr)-\negl(\lambda).
\end{equation}

\emph{Return to the original state.}
The alphabet of $\widehat\Pi$ has size three.  Applying Lemma~\ref{lem:afw} with distance $\epsilon_p$ subtracts
precisely $\kappa_p$.  Conditional entropy of a classical register is nonnegative, so we take the maximum with
zero and optimize over $\mathcal C_\lambda$.  The grid has polynomial size and $2c-1\geq1/\poly(\lambda)$, so the
amplified negligible terms remain negligible uniformly under the optimization.  This
gives~\eqref{eq:single-round-bound}; Appendix~\ref{app:single-round} contains the detailed derivation.
\end{proof}

\section{Multi-Round Composition}\label{sec:main}

We now lift the single-round bound to an arbitrary sequential execution.  The protocol records only the aggregate
conditional test success, so we set
\begin{align}
 \omega&=(1-\beta)\omega_p+\beta\omega_m,\label{eq:aggregate-test-win}\\
 g_\beta(\omega)&=\inf_{(1-\beta)a+\beta b=\omega}g(a,b).
 \label{eq:g-beta}
\end{align}
The infimum accounts for every split between the preimage and equation tests that is compatible with the observed
score.  Let $\underline g_\beta\leq g_\beta$ be a certified convex minorant and choose a globally valid affine
support $h_{\omega_0}(\omega)=a_0\omega+b_0\leq\underline g_\beta(\omega)$ on $[0,1]$.

\paragraph{EAT channels and premises.}
Let $R_{i-1}$ contain the complete joint-device memory before round $i$, including all nominal interfaces and
their communication.  Round $i$ defines the channel
\begin{equation}\label{eq:main-eat-channel}
 \mathcal M_i:R_{i-1}\longrightarrow R_iO_iS_iQ_i,\qquad
 O_i=(\widehat\Pi_i,\widehat M_i),\quad S_i=(K_i,T_i,G_i).
\end{equation}
Here $\widehat\Pi_i\in\{0,1,\perp\}$ is nontrivial for $T_i=0$, while $\widehat M_i\in\{0,1\}$ is nontrivial for
$T_i=1$.  The score $Q_i\in\{\mathsf g,\mathsf g_{\rm bad},0,1\}$ records a valid or invalid generation output, or
a failed or passed test.  Thus, for every attainable score distribution $q$,
\begin{equation}\label{eq:attainable-score-distribution}
 q(\mathsf g)+q(\mathsf g_{\rm bad})=1-\gamma,\qquad
 q(0)+q(1)=\gamma,\qquad \omega=q(1)/\gamma.
\end{equation}

Malformed responses are mapped to rejecting symbols, so $\mathcal M_i$ is trace preserving on every input.  Its
classical output is finite, and $Q_i$ is a deterministic function of $(O_i,T_i,G_i)$.  The side information
$S_i$ is sampled by the verifier independently of everything that precedes it, so the joint state of
$O^{i-1}S^{i-1}ES_i$ is a product with $S_i$ and the Markov condition
\begin{equation}\label{eq:main-eat-markov}
 O^{i-1}\ \longleftrightarrow\ S^{i-1}E\ \longleftrightarrow\ S_i
\end{equation}
holds.  Since $R_{i-1}$ carries every interface, communication register, and coherent memory, this structure
imposes neither an IID nor a non-signaling premise.

It remains to justify applying the computational single-round statement after conditioning on an arbitrary
history.  The conditional state in $R_{i-1}$ need not be efficiently preparable, but it occupies only polynomially
many qubits and is independent of the newly sampled $K_i$.  It is therefore admissible advice in
Definition~\ref{def:ahb-2party}, and Theorem~\ref{thm:single-round} applies to the next response instrument.
Eve's purifying register is retained as side information but never supplied to the computational reduction.

Since $T_i=0$ with probability $1-\beta\gamma$, define
\begin{equation}\label{eq:fmin-def}
 f_{\min}(q)=(1-\beta\gamma)
 \left(a_0\frac{q(1)}\gamma+b_0-\xi_{\rm AHB}(\lambda)\right).
\end{equation}

\begin{lemma}[Affine min-tradeoff function]\label{lem:min-tradeoff}
The function~\eqref{eq:fmin-def} is a min-tradeoff function for the channels~\eqref{eq:main-eat-channel}.
\end{lemma}

\noindent The proof, including the occurrence factor $1-\beta\gamma$ and the conditioning on arbitrary histories,
is given in Appendix~\ref{app:eat-details}.

\paragraph{Accumulation and extraction.}
Let the convex acceptance set be
\begin{equation}\label{eq:accepted-frequency-set}
 \widehat\Omega_0=\left\{q:
 q(1)\geq\gamma\omega_{\rm exp}-\delta_{\rm est},\quad
 q(\mathsf g_{\rm bad})=0,\quad
 q(0)+q(1)\leq\gamma+\delta_{\rm samp}\right\},
\end{equation}
let $\Omega_0$ be the event that the empirical score distribution lies in $\widehat\Omega_0$, and set
\begin{equation}\label{eq:t-definition}
 t=\inf_{q\in\widehat\Omega_0}f_{\min}(q).
\end{equation}
This linear program is evaluated explicitly in Appendix~\ref{app:eat-details}.  If
$\Pr[\Omega_0]\geq\varepsilon_{\rm acc}$, the EAT yields
\begin{equation}\label{eq:eat-complete-output}
 H_{\min}^{\varepsilon_s}(O^n|K^nT^nG^nE)_{\rho|\Omega_0}\geq nt-v\sqrt n,
\end{equation}
where $v$ is evaluated at $\varepsilon_{\rm acc}$ as in Section~\ref{sec:eat-prelim}.

Let $Z_A=(\widehat\Pi_i:G_i=1)$ be Alice's generation string and $O_{\rm test}=(O_i:G_i=0)$ the coarse test
outputs, which determine every announced test result $W_i$.  Conditioning on $O_{\rm test}$, and on the branch
bits and forwarded equation answers announced in test rounds (at most $N$ bits per round), costs, respectively,
\begin{align}
 L_{\rm drop}&\leq(\gamma+\delta_{\rm samp})n\log3,\label{eq:Ldrop}\\
 L_{\rm test}&\leq(\gamma+\delta_{\rm samp})nN.\label{eq:explicit-Ltest}
\end{align}
No other test data become public, so the test leakage does not depend on the length of the preimages or of the
equation strings.  After charging reconciliation and verification by $L_{\rm EC}$, the resulting bound is
\begin{equation}\label{eq:main-final-entropy}
 H_{\min}^{\varepsilon_s}(Z_A|C_{\rm preEC}C_{\rm EC}E)_{\rho|\Omega_0}
 \geq nt-v\sqrt n-L_{\rm drop}-L_{\rm test}-L_{\rm EC},
\end{equation}
where $C_{\rm preEC}$ is the public transcript before reconciliation and $C_{\rm EC}$ the reconciliation and
verification transcript.  Quantum-proof leftover hashing~\cite{Tomamichel2011} and the key-length
condition~\eqref{eq:key-length} then give the secrecy and correctness claims of Theorem~\ref{thm:main}; if
$\Pr[\Omega_0]<\varepsilon_{\rm acc}$, the secrecy criterion holds trivially with error $\varepsilon_{\rm acc}$.
Appendix~\ref{app:eat-details} proves the channel conditions, the fresh-key step, the tradeoff bound, the leakage
accounting, the extraction, and the full theorem.

\section{Everlasting Security}\label{sec:lifting}

The LWE-based AHB property is required only while the device answers the hidden challenges.  Once privacy
amplification has established statistical secrecy, no later computation can weaken it, even if LWE becomes easy.

\begin{proposition}[Post-execution security lifting]\label{prop:lifting}
If a completed execution satisfies
\begin{equation}\label{eq:lift-premise}
 p_\Omega\DeltaTr{\rho_{K_ACE|\Omega}}{\tau_{K_A}\otimes\rho_{CE|\Omega}}
 \leq\varepsilon_{\rm sec},
\end{equation}
then every post-execution channel $\Gamma$ on $K_ACE$, with no restriction on its running time, satisfies
\begin{equation}\label{eq:lift-conclusion}
 p_\Omega\DeltaTr{\Gamma(\rho_{K_ACE|\Omega})}
 {\Gamma(\tau_{K_A}\otimes\rho_{CE|\Omega})}\leq\varepsilon_{\rm sec}.
\end{equation}
\end{proposition}

\begin{proof}
Set $\rho^{\rm acc}=p_\Omega\rho_{K_ACE|\Omega}$ and
$\sigma^{\rm acc}=p_\Omega(\tau_{K_A}\otimes\rho_{CE|\Omega})$.  The premise states that
$\frac12\|\rho^{\rm acc}-\sigma^{\rm acc}\|_1\leq\varepsilon_{\rm sec}$.  Trace distance is contractive under
$\Gamma$, and $\Gamma$ is linear, which gives~\eqref{eq:lift-conclusion}.
\end{proof}

Thus the accepted key remains secure even if an unbounded post-execution algorithm later solves every LWE
instance in the transcript.  LWE hardness is needed until certification and privacy amplification
establish~\eqref{eq:lift-premise}, but not afterward.

\begin{corollary}[Security after a later failure of LWE]\label{cor:lwe-break}
	Assume that LWE is hard during the challenge--response phase, so that Theorem~\ref{thm:main}
	establishes~\eqref{eq:lift-premise}.  After termination, let $\mathcal L$ be any, possibly unbounded,
	algorithm that solves the LWE instances in the public transcript, and let $\mathcal U$ be any key-using
	environment.  Then the real and ideal outputs of
	\begin{equation}
		\Gamma_{\mathcal L,\mathcal U}
		:=\mathcal U\circ(\mathcal L\otimes\operatorname{id})
	\end{equation}
	have abort-weighted trace distance at most $\varepsilon_{\rm sec}$.
\end{corollary}

\begin{proof}
	The composition $\Gamma_{\mathcal L,\mathcal U}$ is a post-execution channel on the systems
	in~\eqref{eq:lift-premise}, irrespective of its running time.  The claim follows from
	Proposition~\ref{prop:lifting}, whose ideal input contains an independent uniform key.
\end{proof}

\paragraph{Comparison with post-quantum group key exchange.}\label{sec:pq-gke-comparison}
Conventional post-quantum group key exchange, such as the Ring-LWE construction of Apon, Dachman-Soled, Gong, and
Katz~\cite{Apon2019}, provides computational indistinguishability against efficient adversaries under a
continuing hardness assumption.  This differs from the statistical conclusion~\eqref{eq:lift-premise}: an
algorithm that later recovers the key with probability $1-\delta$ distinguishes it from uniform with advantage at
least $1-\delta-2^{-\ell}$, which is precisely what~\eqref{eq:lift-conclusion} rules out.

\paragraph{Boundary of the lifting statement.}\label{sec:lifting-boundary}
Proposition~\ref{prop:lifting} concerns the final keys, the public transcript, and Eve's side information.  It
does not cover later leakage of the key or the release of unerased verifier or device state.
Appendix~\ref{app:lifting} gives the full operational statement, its postselection form, and the precise
isolation and erasure requirements.

\section{Honest Noise and Resource Comparison}\label{sec:discussion}

This section evaluates the protocol for a noisy honest implementation: it gives the honest test statistics and
the certified rate, quantifies the noise tolerance, and compares resource costs with pairwise DI-QKD.

For the honest implementation, assume that Alice's local registers are ideal and that each transmitted branch
qubit undergoes independent depolarizing noise
\begin{equation}\label{eq:depolarizing-channel}
 \mathcal D_p(\rho)=(1-p)\rho+p\id/2,\qquad Q=p/2.
\end{equation}
The honest statistics are
\begin{align}
 \omega_p(Q)&=1,\label{eq:omega-p-depol}\\
 \omega_m^{(N)}(Q)&\geq\frac{1+(1-2Q)^{N-1}}2-\nuG(\lambda),\label{eq:omega-m-depol}\\
 \Pr[b_1=\cdots=b_N\mid T=0]&=(1-Q)^{N-1},\label{eq:agreement-depol}
\end{align}
and each Alice--Bob pair has quantum bit error rate $Q$.  As $\gamma,\delta_{\rm samp},\delta_{\rm est}\to0$, and
suppressing the negligible $\nuG$, the aggregate score and the certified asymptotic rate satisfy
\begin{align}
 \omega_N(Q)&=1-\beta+\frac\beta2[1+(1-2Q)^{N-1}],\label{eq:honest-aggregate-score}\\
 r_{N,\infty}^{\rm cert}(Q)&\geq
 [\underline g_\beta(\omega_N(Q))-\hbin(Q)]_+.\label{eq:rate-formula}
\end{align}
This certified rate retains the good-subspace loss, the worst compatible test split, and the convexification.
Appendix~\ref{app:noise} gives the derivation, a finite-noise completeness statement, and low-noise expansions.

\paragraph{Noise tolerance.}
The certified rate is positive only in an extremely narrow window around perfect statistics.  For each $\beta$,
Table~\ref{tab:noise-threshold} lists the aggregate test-failure probability $1-\omega^\ast$ at which the
single-round bound $g_\beta$ of~\eqref{eq:g-beta} becomes zero.  It also lists the depolarizing noise $Q^\ast$ that
produces this failure probability, obtained by solving $\omega_N(Q^\ast)=\omega^\ast$
with~\eqref{eq:honest-aggregate-score}.  Since $\underline g_\beta\leq g_\beta$, the certified
rate~\eqref{eq:rate-formula} is zero for all $Q\geq Q^\ast$.  This is a sufficient condition only.  The rate
already vanishes once $\underline g_\beta(\omega_N(Q))\leq\hbin(Q)$, which may happen at a smaller $Q$, so $Q^\ast$
is an upper bound on the noise that the protocol tolerates, not the exact point at which~\eqref{eq:rate-formula}
becomes zero.  Because $1-\omega_N(Q)=\frac\beta2[1-(1-2Q)^{N-1}]\approx\beta(N-1)Q$, the thresholds $Q^\ast$ scale as
$(N-1)^{-1}$.  At perfect statistics, by contrast, $g_\beta(1)$ is close to one bit.

\paragraph{Numerical evaluation.}
All computations use IEEE double precision.  The maximum over overlaps in~\eqref{eq:mp-entropy} is taken over the
grid $c=\frac12+\delta$, with $6000$ values of $\delta$ spaced logarithmically in $[10^{-4},\frac12]$.  This is a
valid choice of $\mathcal C_\lambda$, so the maximum is computed exactly for this grid.  The infimum
in~\eqref{eq:g-beta} is replaced by a minimum over $601$ test splits.  The preimage failure probability $1-a$
takes the value $0$ and $600$ values spaced logarithmically between $10^{-25}s_{\max}$ and
$s_{\max}=\min\{1,(1-\omega)/(1-\beta)\}$, and the equation failure probability $1-b$ is then fixed by the
constraint.  A minimum over finitely many splits can only overestimate the infimum.  The computed $g_\beta$ is
therefore an upper bound on the exact one, and wherever it is zero, so is the exact $g_\beta$.  In all our runs the
minimum is attained at the endpoint $1-a=s_{\max}$, in which all test failures fall on the preimage test.  The zero
crossing in $1-\omega$ is located by $30$ steps of geometric bisection on $[10^{-14},10^{-10}]$, which brackets it
to a relative width below $10^{-8}$.  Table~\ref{tab:noise-threshold} reports the upper end of the bracket, rounded
to two significant digits.  The tabulated values are thus upper bounds on the zero crossing for the chosen overlap
grid.  Refining either grid by a factor of ten, or extending the overlap grid down to $\delta=10^{-8}$, leaves the
first five significant digits of every threshold unchanged.  The script that produces the table is provided as
an ancillary file (\texttt{anc/noise\_threshold.py}) with the arXiv version of this paper.

\begin{table}[t]
\centering
\begin{tabular}{@{}cccccc@{}}
\toprule
$\beta$ & $1-\omega^\ast$ & $Q^\ast$ ($N=2$) & $Q^\ast$ ($N=3$) & $Q^\ast$ ($N=5$) & $Q^\ast$ ($N=10$)\\
\midrule
$0.5$  & $6.5\times10^{-13}$ & $1.3\times10^{-12}$ & $6.5\times10^{-13}$ & $3.2\times10^{-13}$ & $1.4\times10^{-13}$\\
$0.1$  & $1.2\times10^{-12}$ & $1.2\times10^{-11}$ & $5.8\times10^{-12}$ & $2.9\times10^{-12}$ & $1.3\times10^{-12}$\\
$0.01$ & $1.3\times10^{-12}$ & $1.3\times10^{-10}$ & $6.4\times10^{-11}$ & $3.2\times10^{-11}$ & $1.4\times10^{-11}$\\
\bottomrule
\end{tabular}
\caption{Noise thresholds of the single-round bound.  The function $g_\beta$ vanishes for aggregate test-failure
probabilities $1-\omega\geq1-\omega^\ast$, and $Q^\ast$ is the depolarizing noise at which $1-\omega_N(Q)=1-\omega^\ast$.
The certified rate~\eqref{eq:rate-formula} is therefore zero for $Q\geq Q^\ast$, and it may vanish already at smaller
$Q$.  All values are numerical upper bounds on the zero crossing, rounded to two significant digits (see the
paragraph on numerical evaluation).  The thresholds scale as $(N-1)^{-1}$ and depend only weakly on $\beta$ through
$1-\omega^\ast$.}\label{tab:noise-threshold}
\end{table}

These thresholds are many orders of magnitude below current experimental error rates.  Moreover, near the
threshold any affine minorant of $g_\beta$ that certifies a substantial rate has slope of order at least $10^{11}$,
so the second-order term of the EAT
requires an extremely large number of rounds; cf.\ the finite-size estimates for the same single-round bound
in~\cite{MA2025}.  The limitation stems from the single-round bound, specifically the constant $10/(2c-1)^2$
and the fourth root in $q_c$, and not from the multiparty construction.  Any improvement of that bound applies
directly to Protocol~\ref{prot:dicka}.  

\paragraph{Coherent-evaluation cost.}
A pairwise construction would run $N-1$ computational DI-QKD instances between Alice and the Bobs and use the
resulting pairwise keys to transport a conference key.  Producing one candidate raw symbol for every participant
then requires $N-1$ coherent NTCF evaluations, whereas our multiparty construction uses one evaluation and $N-1$
CNOT gates.  This is a structural saving in the most expensive quantum circuit of the protocol.  We do not claim a
higher key rate: by~\eqref{eq:omega-m-depol} the honest parity success degrades with $N$, and
Table~\ref{tab:noise-threshold} shows that the noise threshold shrinks accordingly.

\section*{Author contributions}
S.C.\ conceived the project, developed the protocol and the security proofs, and wrote the manuscript.
The AI assistant Claude (Anthropic) helped with editing the text, checking notation and cross-references, and
formatting the bibliography.  It also helped write the code for the numerical evaluation in
Section~\ref{sec:discussion}, which produces Table~\ref{tab:noise-threshold}.  The author checked all AI-assisted
text and code and takes full responsibility for the content of the paper.

\bibliographystyle{quantum}
\bibliography{references}

\begin{thebibliography}{10}

\bibitem{BB84}
Charles~H. Bennett and Gilles Brassard.
\newblock ``Quantum cryptography: Public key distribution and coin tossing''.
\newblock In Proceedings of the IEEE International Conference on Computers,
  Systems and Signal Processing.
\newblock \href{https://dx.doi.org/10.1016/j.tcs.2014.05.025}{Pages 175--179}.
\newblock ~(1984).

\bibitem{Ekert1991}
Artur~K. Ekert.
\newblock ``Quantum cryptography based on {Bell}'s theorem''.
\newblock \href{https://dx.doi.org/10.1103/PhysRevLett.67.661}{Physical Review
  Letters {\bf 67}, 661--663}~(1991).

\bibitem{Lydersen2010}
Lars Lydersen, Carlos Wiechers, Christoffer Wittmann, Dominique Elser, Johannes
  Skaar, and Vadim Makarov.
\newblock ``Hacking commercial quantum cryptography systems by tailored bright
  illumination''.
\newblock \href{https://dx.doi.org/10.1038/nphoton.2010.214}{Nature Photonics
  {\bf 4}, 686--689}~(2010).

\bibitem{Xu2020}
Feihu Xu, Xiongfeng Ma, Qiang Zhang, Hoi-Kwong Lo, and Jian-Wei Pan.
\newblock ``Secure quantum key distribution with realistic devices''.
\newblock \href{https://dx.doi.org/10.1103/RevModPhys.92.025002}{Reviews of
  Modern Physics {\bf 92}, 025002}~(2020).

\bibitem{MY1998}
Dominic Mayers and Andrew Yao.
\newblock ``Quantum cryptography with imperfect apparatus''.
\newblock In 39th Annual Symposium on Foundations of Computer Science.
\newblock \href{https://dx.doi.org/10.1109/SFCS.1998.743501}{Pages 503--509}.
\newblock IEEE~(1998).

\bibitem{ABG2007}
Antonio Ac{\'{i}}n, Nicolas Brunner, Nicolas Gisin, Serge Massar, Stefano
  Pironio, and Valerio Scarani.
\newblock ``Device-independent security of quantum cryptography against
  collective attacks''.
\newblock \href{https://dx.doi.org/10.1103/PhysRevLett.98.230501}{Physical
  Review Letters {\bf 98}, 230501}~(2007).

\bibitem{VV2014}
Umesh Vazirani and Thomas Vidick.
\newblock ``Fully device-independent quantum key distribution''.
\newblock \href{https://dx.doi.org/10.1103/PhysRevLett.113.140501}{Physical
  Review Letters {\bf 113}, 140501}~(2014).

\bibitem{Hensen2015}
Bas Hensen et~al.
\newblock ``Loophole-free {Bell} inequality violation using electron spins
  separated by 1.3 kilometres''.
\newblock \href{https://dx.doi.org/10.1038/nature15759}{Nature {\bf 526},
  682--686}~(2015).

\bibitem{Giustina2015}
Marissa Giustina et~al.
\newblock ``Significant-loophole-free test of {Bell}'s theorem with entangled
  photons''.
\newblock \href{https://dx.doi.org/10.1103/PhysRevLett.115.250401}{Physical
  Review Letters {\bf 115}, 250401}~(2015).

\bibitem{Shalm2015}
Lynden~K. Shalm et~al.
\newblock ``Strong loophole-free test of local realism''.
\newblock \href{https://dx.doi.org/10.1103/PhysRevLett.115.250402}{Physical
  Review Letters {\bf 115}, 250402}~(2015).

\bibitem{Nadlinger2022}
David~P. Nadlinger, Peter Drmota, Bethan~C. Nichol, Gabriel Araneda, David
  Main, Raghavendra Srinivas, David~M. Lucas, Christopher~J. Ballance, Kirill
  Ivanov, Ernest Y.-Z. Tan, Pavel Sekatski, R{\"u}diger~L. Urbanke, Renato
  Renner, Nicolas Sangouard, and Jean-Daniel Bancal.
\newblock ``Experimental quantum key distribution certified by {Bell}'s
  theorem''.
\newblock \href{https://dx.doi.org/10.1038/s41586-022-04941-5}{Nature {\bf
  607}, 682--686}~(2022).

\bibitem{Zhang2022}
Wei Zhang et~al.
\newblock ``A device-independent quantum key distribution system for distant
  users''.
\newblock \href{https://dx.doi.org/10.1038/s41586-022-04891-y}{Nature {\bf
  607}, 687--691}~(2022).

\bibitem{MurtaReview2020}
Gl{\'{a}}ucia Murta, Federico Grasselli, Hermann Kampermann, and Dagmar
  Bru{\ss}.
\newblock ``Quantum conference key agreement: A review''.
\newblock \href{https://dx.doi.org/10.1002/qute.202000025}{Advanced Quantum
  Technologies {\bf 3}, 2000025}~(2020).

\bibitem{Ribeiro2018}
Jeremy Ribeiro, Gl{\'{a}}ucia Murta, and Stephanie Wehner.
\newblock ``Fully device-independent conference key agreement''.
\newblock \href{https://dx.doi.org/10.1103/PhysRevA.97.022307}{Physical Review
  A {\bf 97}, 022307}~(2018).

\bibitem{Holz2020}
Timo Holz, Hermann Kampermann, and Dagmar Bru{\ss}.
\newblock ``Genuine multipartite {Bell} inequality for device-independent
  conference key agreement''.
\newblock \href{https://dx.doi.org/10.1103/PhysRevResearch.2.023251}{Physical
  Review Research {\bf 2}, 023251}~(2020).

\bibitem{BCMVV2018}
Zvika Brakerski, Paul Christiano, Urmila Mahadev, Umesh Vazirani, and Thomas
  Vidick.
\newblock ``A cryptographic test of quantumness and certifiable randomness from
  a single quantum device''.
\newblock In 59th Annual Symposium on Foundations of Computer Science.
\newblock \href{https://dx.doi.org/10.1109/FOCS.2018.00038}{Pages 320--331}.
\newblock IEEE~(2018).

\bibitem{Mahadev2018Verification}
Urmila Mahadev.
\newblock ``Classical verification of quantum computations''.
\newblock In 59th Annual Symposium on Foundations of Computer Science.
\newblock \href{https://dx.doi.org/10.1109/FOCS.2018.00033}{Pages 259--267}.
\newblock IEEE~(2018).

\bibitem{GheorghiuVidick2019}
Alexandru Gheorghiu and Thomas Vidick.
\newblock ``Computationally-secure and composable remote state preparation''.
\newblock In 60th Annual Symposium on Foundations of Computer Science.
\newblock \href{https://dx.doi.org/10.1109/FOCS.2019.00066}{Pages 1024--1033}.
\newblock IEEE~(2019).

\bibitem{BartusekKhurana2025}
James Bartusek and Dakshita Khurana.
\newblock ``On the power of oblivious state preparation''.
\newblock In Advances in Cryptology -- CRYPTO 2025.
\newblock \href{https://dx.doi.org/10.1007/978-3-032-01878-6_19}{Pages
  575--607}.
\newblock Lecture Notes in Computer Science. Springer~(2025).

\bibitem{KLVY2023}
Yael~Tauman Kalai, Alex Lombardi, Vinod Vaikuntanathan, and Lisa Yang.
\newblock ``Quantum advantage from any non-local game''.
\newblock In 55th Annual ACM Symposium on Theory of Computing.
\newblock \href{https://dx.doi.org/10.1145/3564246.3585162}{Pages 1617--1628}.
\newblock ACM~(2023).

\bibitem{Metger2024}
Tony Metger, Yfke Dulek, Andrea Coladangelo, and Rotem Arnon-Friedman.
\newblock ``Device-independent quantum key distribution from computational
  assumptions''.
\newblock \href{https://dx.doi.org/10.1088/1367-2630/ac304b}{New Journal of
  Physics {\bf 23}, 123021}~(2021).

\bibitem{MA2025}
Ilya Merkulov and Rotem Arnon.
\newblock ``Entropy accumulation under post-quantum cryptographic
  assumptions''.
\newblock \href{https://dx.doi.org/10.3390/e27080772}{Entropy {\bf 27},
  772}~(2025).

\bibitem{TomamichelRenner2011}
Marco Tomamichel and Renato Renner.
\newblock ``Uncertainty relation for smooth entropies''.
\newblock \href{https://dx.doi.org/10.1103/PhysRevLett.106.110506}{Physical
  Review Letters {\bf 106}, 110506}~(2011).

\bibitem{DFR2020}
Fr{\'{e}}d{\'{e}}ric Dupuis, Omar Fawzi, and Renato Renner.
\newblock ``Entropy accumulation''.
\newblock \href{https://dx.doi.org/10.1007/s00220-020-03839-5}{Communications
  in Mathematical Physics {\bf 379}, 867--913}~(2020).

\bibitem{MFSR2022}
Tony Metger, Omar Fawzi, David Sutter, and Renato Renner.
\newblock ``Generalised entropy accumulation''.
\newblock In 63rd Annual Symposium on Foundations of Computer Science.
\newblock \href{https://dx.doi.org/10.1109/FOCS54457.2022.00085}{Pages
  844--850}.
\newblock IEEE~(2022).

\bibitem{AFDFRV2018}
Rotem Arnon-Friedman, Fr{\'{e}}d{\'{e}}ric Dupuis, Omar Fawzi, Renato Renner,
  and Thomas Vidick.
\newblock ``Practical device-independent quantum cryptography via entropy
  accumulation''.
\newblock \href{https://dx.doi.org/10.1038/s41467-017-02307-4}{Nature
  Communications {\bf 9}, 459}~(2018).

\bibitem{Grasselli2019}
Federico Grasselli, Hermann Kampermann, and Dagmar Bru{\ss}.
\newblock ``Conference key agreement with single-photon interference''.
\newblock \href{https://dx.doi.org/10.1088/1367-2630/ab573e}{New Journal of
  Physics {\bf 21}, 123002}~(2019).

\bibitem{Baroni2025}
Matilde Baroni, Igor Klep, Dominik Leichtle, Marc-Olivier Renou, Ivan
  {\v{S}}upi{\'c}, Lucas Tendick, and Xiangling Xu.
\newblock ``Quantitative quantum soundness for all multipartite compiled
  nonlocal games''~(2025).
\newblock  \href{http://arxiv.org/abs/2509.25145}{arXiv:2509.25145}.

\bibitem{AH2023}
Scott Aaronson and Shih-Han Hung.
\newblock ``Certified randomness from quantum supremacy''.
\newblock In 55th Annual ACM Symposium on Theory of Computing.
\newblock \href{https://dx.doi.org/10.1145/3564246.3585145}{Pages 933--944}.
\newblock ACM~(2023).

\bibitem{Apon2019}
Daniel Apon, Dana Dachman-Soled, Huijing Gong, and Jonathan Katz.
\newblock ``Constant-round group key exchange from the {Ring-LWE} assumption''.
\newblock In Post-Quantum Cryptography.
\newblock \href{https://dx.doi.org/10.1007/978-3-030-25510-7_11}{Volume 11505
  of Lecture Notes in Computer Science, pages 189--205}.
\newblock Springer~(2019).

\bibitem{Renner2008}
Renato Renner.
\newblock ``Security of quantum key distribution''.
\newblock \href{https://dx.doi.org/10.1142/S0219749908003256}{International
  Journal of Quantum Information {\bf 6}, 1--127}~(2008).

\bibitem{BenOr2005}
Michael Ben-Or, Micha{\l} Horodecki, Debbie~W. Leung, Dominic Mayers, and
  Jonathan Oppenheim.
\newblock ``The universal composable security of quantum key distribution''.
\newblock In Theory of Cryptography Conference.
\newblock \href{https://dx.doi.org/10.1007/978-3-540-30576-7_21}{Volume 3378 of
  Lecture Notes in Computer Science, pages 386--406}.
\newblock Springer~(2005).

\bibitem{Canetti2001}
Ran Canetti.
\newblock ``Universally composable security: A new paradigm for cryptographic
  protocols''.
\newblock In 42nd Annual Symposium on Foundations of Computer Science.
\newblock \href{https://dx.doi.org/10.1109/SFCS.2001.959888}{Pages 136--145}.
\newblock IEEE~(2001).

\bibitem{Winter1999}
Andreas Winter.
\newblock ``Coding theorem and strong converse for quantum channels''.
\newblock \href{https://dx.doi.org/10.1109/18.796385}{IEEE Transactions on
  Information Theory {\bf 45}, 2481--2485}~(1999).

\bibitem{Ogawa2005}
Tomohiro Ogawa and Hiroshi Nagaoka.
\newblock ``Making good codes for classical-quantum channel coding via quantum
  hypothesis testing''.
\newblock \href{https://dx.doi.org/10.1109/TIT.2007.896874}{IEEE Transactions
  on Information Theory {\bf 53}, 2261--2266}~(2007).

\bibitem{AlickiFannes2004}
Robert Alicki and Mark Fannes.
\newblock ``Continuity of quantum conditional information''.
\newblock \href{https://dx.doi.org/10.1088/0305-4470/37/5/L01}{Journal of
  Physics A: Mathematical and General {\bf 37}, L55--L57}~(2004).

\bibitem{Winter2016}
Andreas Winter.
\newblock ``Tight uniform continuity bounds for quantum entropies: Conditional
  entropy, relative entropy distance and energy constraints''.
\newblock \href{https://dx.doi.org/10.1007/s00220-016-2609-8}{Communications in
  Mathematical Physics {\bf 347}, 291--313}~(2016).

\bibitem{Tomamichel2011}
Marco Tomamichel, Christian Schaffner, Adam Smith, and Renato Renner.
\newblock ``Leftover hashing against quantum side information''.
\newblock \href{https://dx.doi.org/10.1109/TIT.2011.2158473}{IEEE Transactions
  on Information Theory {\bf 57}, 5524--5535}~(2011).

\end{thebibliography}

\appendix
\section{Detailed Collective-AHB Interface}\label{app:collective-ahb}

This appendix details the interface between the collective AHB property (Section~\ref{sec:mp-ahb}) and the
operator argument (Section~\ref{sec:mp-goodsub}).  It explains how the reduction implements the analytical
sequential experiment, why Alice's preimage answer must be measured without the Bobs' answers, and how the event
probabilities become the bias of the balanced binary measurement.  Only registers available to the efficient device
occur in the computational reduction; Eve's purifying system enters later, through the entropic uncertainty
relation.

\subsection{Implementing Alice's marginal measurement}

Fix a fresh key $k$, and let $\varphi^{k,y}$ be the subnormalized device state after the device has committed to
$y$.  The device's preimage-response instrument is implemented by an efficient unitary $U_y$ followed by a
measurement of output registers holding Alice's answer $(b,x)$ and the Bobs' answers $\vec b$.  Writing the
dilated projectors as
\begin{equation}
 P_y^{b,x,\vec b}=U_y^\dagger\bigl(|b,x\rangle\!\langle b,x|\otimes|\vec b\rangle\!\langle\vec b|\otimes\id\bigr)U_y,
\end{equation}
the projector~\eqref{eq:alice-marginal} is
$\Pi_y^{b,x}=U_y^\dagger(|b,x\rangle\!\langle b,x|\otimes\id)U_y$.  It is implemented by applying $U_y$, measuring
only Alice's output registers, and applying $U_y^\dagger$.  The equation-response instrument is dilated in the same
way with fresh work registers, so that both measurements act on a common space.  All of these steps are efficient,
so the reduction $\cA^*$ of Theorem~\ref{thm:mp-ahb} can run the sequential experiment.  Whether $y$ is well formed
cannot be decided without the trapdoor, but this is never needed: the reduction only tests the validity of Alice's
answer with $\Chk$, and a valid answer certifies that $y$ is well formed (Lemma~\ref{lem:well-formed}).

Measuring the Bobs' preimage answers as well would produce the fine-grained experiment, whose probabilities
\begin{equation}
 \sum_{\vec b}\Tr\!\left(M_y^rP_y^{b,x,\vec b}\varphi^{k,y}P_y^{b,x,\vec b}\right)
 \quad\text{differ in general from}\quad
 \Tr\!\left(M_y^r\Pi_y^{b,x}\varphi^{k,y}\Pi_y^{b,x}\right),
\end{equation}
because the latter retains the cross terms between different values
of $\vec b$.  The operator argument uses the projectors $\PiN{b}(y)$ of~\eqref{eq:preimage-projectors}, which
retain these cross terms, so the reduction must use Alice's marginal measurement.  In the two-party setting
of~\cite{BCMVV2018,MA2025} the issue does not arise: there are no Bobs, and a valid preimage is unique given
$(b,y)$.

\subsection{Event probabilities as traces}

For a well-formed image $y$ and a valid answer $(b,x_b(y))$, the state after the preimage measurement
is $\PiN{b}(y)\varphi^{k,y}\PiN{b}(y)$.  Applying the equation measurement and summing over the classes of
Definition~\ref{def:mp-ahb} gives
\begin{equation}\label{eq:trace-event-identity}
\begin{aligned}
 \Pr[\sC]&=\E_k\sum_{y,b}\Tr\!\left(M_{\sC}(y)\PiN{b}(y)\varphi^{k,y}\PiN{b}(y)\right),\\
 \Pr[\sI]&=\E_k\sum_{y,b}\Tr\!\left(M_{\sI}(y)\PiN{b}(y)\varphi^{k,y}\PiN{b}(y)\right),
\end{aligned}
\end{equation}
which is~\eqref{eq:ahb-trace-form}.  The deterministic parity map~\eqref{eq:transcript-map} sends $\sC$ and $\sI$
to the ordinary AHB correct and incorrect events.  On every other outcome the reduction emits a well-typed prefix
and an independent fair bit; conditioned on any fixed prefix, this contributes equally to the two ordinary events.
Therefore
\begin{equation}\label{eq:exact-advantage-identity}
 \left|\Pr[\sC]-\Pr[\sI]\right|
 =\bigl|\Pr[\cA^*\in H_k]-\Pr[\cA^*\in\bar H_k]\bigr|
 \le\eta_{\rm AHB}(\lambda).
\end{equation}
No branch-agreement predicate appears in~\eqref{eq:trace-event-identity}.  Such a predicate would remove only some
valid transcripts and would have no exact counterpart in the ordinary AHB game.

\subsection{From AHB events to the binary projector bias}

Condition on a valid preimage and let $\Psi$ be the normalized state in~\eqref{eq:valid-projected-state}.  Pinch it
in the binary preimage measurement:
\begin{equation}
 \mathcal P_{\Pi}(\Psi)
 :=\PiN{0}\Psi\PiN{0}+\PiN{1}\Psi\PiN{1}.
\end{equation}
Let $p_\sC=\Pr[\sC]/\omega_p$, $p_\sI=\Pr[\sI]/\omega_p$, and $p_\sB=1-p_\sC-p_\sI$ be the conditional probabilities
that the subsequent equation answer is correct-admissible, incorrect-admissible, or in $\sB$.  On the enlarged state
with the independent uniform coin $\Coin$, the balanced measurement obeys
\begin{align}
 \Tr(\MBN{0}\mathcal P_{\Pi}(\Psi))&=p_\sC+\tfrac12p_\sB,\\
 \Tr(\MBN{1}\mathcal P_{\Pi}(\Psi))&=p_\sI+\tfrac12p_\sB.
\end{align}
Consequently the bias used in the good-subspace lemma is
\begin{equation}\label{eq:balanced-mu-identity}
 \mu
 =\left|\frac12-
 \Tr(\MBN{0}\PiN{0}\Psi\PiN{0})-
 \Tr(\MBN{0}\PiN{1}\Psi\PiN{1})\right|
 =\frac12|p_\sC-p_\sI|
 \leq\frac{\eta_{\rm AHB}(\lambda)}{2\omega_p},
\end{equation}
by~\eqref{eq:exact-advantage-identity}.  This is Lemma~\ref{lem:balanced-completion}.  The equal split of the class
$\sB$ is what makes the identity exact while the protocol retains operational rejection of every inadmissible
answer.

\subsection{Conditioning on earlier rounds}

For round $i$, the state entering the calculation above may depend arbitrarily on the public transcript and device
memory from rounds $1,\ldots,i-1$, and it may be inefficient to prepare.  Its number of qubits is polynomial,
because the device is a polynomial-size machine.  Since $(k_i,\td_i)$ is sampled only after this state exists, the
state is independent of $k_i$.  It is therefore a valid advice state in Definition~\ref{def:ahb-2party}.  This is
exactly why the one-round bound is uniform over the conditional states required by entropy accumulation.

\section{Proof of the Good-Subspace Projection Lemma}\label{app:goodsub-proof}

For completeness, we reproduce the operator argument underlying the good-subspace lemma.  It specializes the
projection lemma used in~\cite{BCMVV2018,MA2025} to the global device.  Since only two projections occur, the proof
is insensitive to any internal tensor-product structure.

Let $\{B_j\}_j$ be the Jordan-block projectors for $\Pi$ and $M$.  One-dimensional blocks can be padded to dimension
two.  On a two-dimensional block choose a basis in which
\begin{equation}\label{eq:appendix-jordan-form}
 \Pi_j=\begin{pmatrix}1&0\\0&0\end{pmatrix},\qquad
 M_j=\begin{pmatrix}a_j^2&a_jb_j\\a_jb_j&b_j^2\end{pmatrix},
\end{equation}
where $a_j=\cos\theta_j$, $b_j=\sin\theta_j$, and
\begin{equation}
 c_j:=\max\{a_j^2,b_j^2\}\in[1/2,1].
\end{equation}
The good-subspace projector is
\begin{equation}
 \Gamma(c)=\sum_{j:c_j\le c}B_j.
\end{equation}

\paragraph{Perfect equation success.}
First suppose $\omega=\Tr(M\psi)=1$.  Then $\psi$ is supported in the range of $M$.  Let $p_j=\Tr(B_j\psi)$.  Because
$M$ has rank one on every nontrivial block, the diagonal block $B_j\psi B_j$ equals $p_jM_j$.  Cross-block coherences
do not contribute to the traces of the block-diagonal operators below.  Direct multiplication
using~\eqref{eq:appendix-jordan-form} gives
\begin{align}
 \Tr(M\Pi\psi\Pi)&=\sum_jp_ja_j^4,\\
 \Tr(M(\id-\Pi)\psi(\id-\Pi))&=\sum_jp_jb_j^4.
\end{align}
Since $a_j^2+b_j^2=1$,
\begin{equation}
 a_j^4+b_j^4
 =1-2a_j^2b_j^2
 =\frac12+\frac12(2c_j-1)^2.
\end{equation}
The quantity inside the absolute value in~\eqref{eq:goodsub-mu} is nonpositive, and therefore
\begin{equation}\label{eq:mu-block-sum}
 2\mu=\sum_jp_j(2c_j-1)^2.
\end{equation}
Every bad block has $c_j>c$, so~\eqref{eq:mu-block-sum} implies
\begin{equation}\label{eq:perfect-goodsub-bound}
 2\mu\ge(2c-1)^2\sum_{j:c_j>c}p_j
 =(2c-1)^2\Tr((\id-\Gamma(c))\psi).
\end{equation}
This proves~\eqref{eq:goodsub-exact} when $\omega=1$, with the stronger numerator $2\mu$.

\paragraph{Imperfect equation success.}
If $\omega=0$, the claimed inequality is immediate, because its right-hand side is at least ten while the left-hand
side is at most one.  Assume $\omega>0$ and set
\begin{equation}
 \psi':=\frac{M\psi M}{\Tr(M\psi)}.
\end{equation}
Then $\psi'$ is supported in the range of $M$, and the gentle projection lemma gives
\begin{equation}\label{eq:appendix-gentle}
 \lVert\psi'-\psi\rVert_1\le2\sqrt{1-\omega}.
\end{equation}
Let $\mu'$ be~\eqref{eq:goodsub-mu} evaluated on $\psi'$.  Bounding the changes in its two trace terms separately and
using that all projections have operator norm one gives the conservative estimate
\begin{equation}\label{eq:mu-prime-bound}
 \mu'\le\mu+4\sqrt{1-\omega}.
\end{equation}
Applying the perfect-success result~\eqref{eq:perfect-goodsub-bound} to $\psi'$ yields
\begin{equation}
 \Tr((\id-\Gamma(c))\psi')
 \le\frac{2\mu+8\sqrt{1-\omega}}{(2c-1)^2}.
\end{equation}
Finally,~\eqref{eq:appendix-gentle} and H\"older duality imply
\begin{align}
 \Tr((\id-\Gamma(c))\psi)
 &\le\Tr((\id-\Gamma(c))\psi')+2\sqrt{1-\omega}\\
 &\le\frac{2\mu+8\sqrt{1-\omega}}{(2c-1)^2}
      +2\sqrt{1-\omega}\\
 &\le\frac{2\mu+10\sqrt{1-\omega}}{(2c-1)^2},
\end{align}
where the last step uses $(2c-1)^2\le1$.  This is~\eqref{eq:goodsub-exact}.

\section{Detailed Derivation of the Single-Round Entropy Bound}\label{app:single-round}

This appendix expands the proof of Theorem~\ref{thm:single-round}.  It makes explicit the two state projections,
the origin of the fourth root, the role of the key and image registers, and the monotonicity domain of the binary
entropy.  The derivation follows~\cite[Lemma~31]{MA2025}, with the collective reduction and the balanced completion
supplying its computational premise.

\subsection{The controlled cq state}

Write the normalized state after a fresh key and image have been produced as
\begin{equation}\label{eq:appendix-cq-state}
 \Phi_{KYDE}=\sum_{k,y}p(k,y)|k,y\rangle\!\langle k,y|_{KY}
 \otimes\Phi^{k,y}_{DE}.
\end{equation}
Define global projectors controlled by the classical registers:
\begin{equation}
 P_Y:=\PiN{0}(Y)+\PiN{1}(Y),\qquad M_Y:=\MBN{0}(Y).
\end{equation}
The valid-preimage probability and the operational equation-test success are
\begin{equation}
 \omega_p=\Tr(P_Y\Phi),\qquad
 \omega_m=\Tr(\MN{0}(Y)\Phi).
\end{equation}
Write $\bar\omega_m=\Tr(M_Y(\Phi\otimes\tau_{\Coin}))$ for the balanced success.  By
Lemma~\ref{lem:balanced-completion}, $\bar\omega_m\ge\omega_m$.  We suppress the coin register below.  If
$g(\omega_p,\omega_m)=0$ there is nothing to prove, so assume $g>0$; then $1-\omega_p<20^{-4}$.  Project onto a valid
Alice preimage and normalize:
\begin{equation}\label{eq:appendix-Psi}
 \Psi=\frac{P_Y\Phi P_Y}{\omega_p}.
\end{equation}
The gentle projection lemma gives
\begin{equation}\label{eq:appendix-valid-distance}
 \frac12\|\Phi-\Psi\|_1\le\sqrt{1-\omega_p}=:\epsilon_p.
\end{equation}
Working with the controlled global projectors is essential.  The AHB property bounds an experiment averaged over the
fresh key and image; it need not provide a pointwise bound for every fixed $(k,y)$.

\subsection{Equation behavior after the valid projection}

Let $M_Y^\perp=\id-M_Y$.  From~\eqref{eq:appendix-valid-distance} and
$|\Tr(A(\rho-\sigma))|\le\|A\|_\infty\|\rho-\sigma\|_1$,
\begin{align}
 \Tr(M_Y^\perp\Psi)
 &\le \Tr(M_Y^\perp\Phi)+\|\Psi-\Phi\|_1\nonumber\\
 &\le 1-\bar\omega_m+2\sqrt{1-\omega_p}\nonumber\\
 &\le 1-\omega_m+2\sqrt{1-\omega_p}.
 \label{eq:appendix-equation-failure}
\end{align}
Taking a square root and using $\sqrt{r+s}\le\sqrt r+\sqrt s$ gives
\begin{align}
 \sqrt{\Tr(M_Y^\perp\Psi)}
 &\le\sqrt{1-\omega_m}
 +\sqrt2\,(1-\omega_p)^{1/4}\nonumber\\
 &\le\sqrt{1-\omega_m}+2(1-\omega_p)^{1/4}.
 \label{eq:appendix-fourth-root}
\end{align}
The same trace-distance estimate in the opposite direction gives
\begin{equation}\label{eq:appendix-equation-success}
 \Tr(M_Y\Psi)\ge\bar\omega_m-2\sqrt{1-\omega_p}
 \ge\omega_m-2\sqrt{1-\omega_p}.
\end{equation}
Equation~\eqref{eq:appendix-fourth-root} exhibits the nested square roots responsible for the fourth-root preimage
penalty.

\subsection{AHB bias and the bad-block weight}

On the support of $P_Y$, take $\Pi_Y=\PiN{0}(Y)$, so that $P_Y-\Pi_Y=\PiN{1}(Y)$, and define
\begin{equation}\label{eq:appendix-global-mu}
 \mu=\left|\frac12
 -\Tr(M_Y\Pi_Y\Psi\Pi_Y)
 -\Tr(M_Y(P_Y-\Pi_Y)\Psi(P_Y-\Pi_Y))\right|.
\end{equation}
By Lemma~\ref{lem:balanced-completion}, with the key and image retained as classical registers,
$\mu\le\eta_{\rm AHB}(\lambda)/(2\omega_p)\le\eta_{\rm AHB}(\lambda)$, since $\omega_p\geq1/2$.

Jordan's lemma may be applied to the controlled global projectors.  Since the projectors onto classical values of
$(k,y)$ commute with both of them, the Jordan decomposition can simultaneously refine those values.  Let
$\Gamma_Y(c)$ be the sum of the blocks of square overlap at most $c$.  The good-subspace lemma
and~\eqref{eq:appendix-fourth-root} yield
\begin{align}
 \Tr((\id-\Gamma_Y(c))\Psi)
 &\le A(c)\left(\frac{\mu}{5}
      +\sqrt{\Tr(M_Y^\perp\Psi)}\right)\nonumber\\
 &\le A(c)\left(2(1-\omega_p)^{1/4}
      +\sqrt{1-\omega_m}\right)+\negl(\lambda)\nonumber\\
 &=q_c+\negl(\lambda),
 \label{eq:appendix-qc}
\end{align}
where the first line is~\eqref{eq:goodsub-exact} rewritten with $A(c)=10/(2c-1)^2$, and the second uses
$A(c)\leq\poly(\lambda)$ on the grid $\mathcal C_\lambda$.

\subsection{Uncertainty inside the good subspace}

The block projector $\Gamma_Y(c)$ commutes with both binary measurements and may be measured without changing their
outcome distributions.  On a good block their square overlap is at most $c$.  The tripartite uncertainty
relation~\eqref{eq:eur}, applied with any system $\tilde{E}$, gives
\begin{equation}\label{eq:appendix-eur-good}
 H(\widehat\Pi|EYK,\Gamma=\mathsf{good})_\Psi
 \ge\log(1/c)
 -H(\widehat M|\tilde{E}YK,\Gamma=\mathsf{good})_\Psi,
\end{equation}
and conditioning does not increase entropy, so the last term is at most $H(\widehat M|\Gamma=\mathsf{good})_\Psi$.

For arbitrary events $A,B$, $\Pr[A\cap B]\ge\Pr[A]+\Pr[B]-1$.
Using~\eqref{eq:appendix-equation-success} and~\eqref{eq:appendix-qc}, and noting that division by $\Pr[B]\le1$ cannot
weaken a nonnegative lower bound, gives
\begin{equation}\label{eq:appendix-zc}
 \Pr[\widehat M=0\mid\Gamma=\mathsf{good}]
 \ge\omega_m-2\sqrt{1-\omega_p}-q_c-\negl(\lambda)
 =z_c-\negl(\lambda).
\end{equation}
When $z_c\ge1/2$, the binary entropy is decreasing on $[1/2,1]$, so
\begin{equation}
 H(\widehat M|\Gamma=\mathsf{good})_\Psi
 \le\hbin(z_c)+\negl(\lambda).
\end{equation}
The conditional entropy of a classical register given a quantum register is nonnegative.  We may therefore condition
on the block flag and discard the bad-block contribution:
\begin{align}
 H(\widehat\Pi|EYK)_\Psi
 &\ge H(\widehat\Pi|EYK\Gamma)_\Psi\\
 &\ge(1-q_c)_+
 \left(\log(1/c)-\hbin(z_c)\right)-\negl(\lambda).
 \label{eq:appendix-entropy-Psi}
\end{align}

\subsection{Continuity and uniform optimization}\label{app:uniform-opt}

The output alphabet $\{0,1,\perp\}$ has size three.  Applying Lemma~\ref{lem:afw}
to~\eqref{eq:appendix-valid-distance} gives
\begin{equation}\label{eq:appendix-continuity}
 H(\widehat\Pi|EYK)_\Phi
 \ge H(\widehat\Pi|EYK)_\Psi
 -\epsilon_p\log3
 -(1+\epsilon_p)\hbin\!\left(\frac{\epsilon_p}{1+\epsilon_p}\right).
\end{equation}
Combining~\eqref{eq:appendix-entropy-Psi} and~\eqref{eq:appendix-continuity} gives the expression inside the
maximization in~\eqref{eq:mp-entropy}.  The restriction $z_c\in[1/2,1]$ records exactly the interval on which the
monotonicity of the binary entropy was used.

For each fixed admissible $c$, the computational correction is negligible.  Optimization over an arbitrary continuum
could hide a nonuniform choice and a multiplication by an uncontrolled $A(c)=10/(2c-1)^2$.  Restricting to a publicly
declared polynomial-size grid with $2c-1\ge1/\poly(\lambda)$ permits a union bound and keeps every amplified
negligible term negligible.  This establishes the uniform $\xi_{\rm AHB}(\lambda)$ in Theorem~\ref{thm:single-round}.

\section{EAT Channels, Leakage, and Extraction Details}\label{app:eat-details}

This appendix verifies the channel conditions and the passage from the complete EAT output to an abort-weighted
conference key.  It also proves that the single-round theorem remains applicable after arbitrary histories, and it
gives the complete proof of Theorem~\ref{thm:main}.

\subsection{Construction of the round channels}

At the beginning of round $i$, $R_{i-1}$ contains the entire joint device memory, including every nominal component
and all communication registers retained from earlier rounds.  A reference $R'$ may purify its input.  The verifier
independently samples $(K_i,T_i,G_i)$ with the joint law specified by Protocol~\ref{prot:dicka}: $G_i$ is hidden,
$T_i=0$ whenever $G_i=1$, and conditional on $G_i=0$, $T_i$ is Bernoulli with parameter $\beta$.  The public key is
fresh.

The round interaction and its classical coarse graining define a channel
\begin{equation}\label{eq:appendix-eat-channel}
 \mathcal M_i:R_{i-1}\longrightarrow
 R_i\,\widehat\Pi_i\,\widehat M_i\,K_iT_iG_iQ_i.
\end{equation}
The map includes key generation, the device's image and response instruments, trapdoor verification, and the
classical computation of the score.  Malformed images are mapped to rejecting outcomes, so the channel is defined on
every input.

\begin{lemma}[EAT channel conditions]\label{lem:appendix-eat-channels}
The maps~\eqref{eq:appendix-eat-channel} satisfy the finite-output condition, the score-reconstruction
condition, and the Markov condition required by the EAT.
\end{lemma}

\begin{proof}
The coarse output alphabet is a subset of $\{0,1,\perp\}\times\{0,1\}$, so $d_O\le6$.  The score is obtained
deterministically from $(\widehat\Pi_i,\widehat M_i,T_i,G_i)$: the generation symbols record whether $\widehat\Pi_i$
is valid, and the test symbols record the accepting bit of the selected test.  This proves score reconstruction.

For the Markov condition, the verifier samples $S_i=(K_i,T_i,G_i)$ with fresh randomness, independently of all
registers that exist before round $i$.  Hence the reduced state on $O^{i-1}S^{i-1}ES_i$ is a product
$\rho_{O^{i-1}S^{i-1}E}\otimes\rho_{S_i}$, and the chain
\begin{equation}
 O^{i-1}\ \longleftrightarrow\ S^{i-1}E\ \longleftrightarrow\ S_i
\end{equation}
holds trivially.  The later use of $T_i$ by the device occurs inside $\mathcal M_i$ and does not affect this
input-side independence.
\end{proof}

\subsection{Fresh keys after arbitrary histories}

\begin{lemma}[Fresh-key conditioning]\label{lem:fresh-key-history}
Fix a round $i$ and a classical history before $K_i$ is sampled.  The conditional state of the carried device memory
is admissible advice for the AHB reduction, and Theorem~\ref{thm:single-round} applies to the next response
instrument.
\end{lemma}

\begin{proof}
The register $R_{i-1}$ contains all earlier behavior, including the communication among the nominal interfaces and
any coherently retained information.  Conditioning on the classical history can leave a state that is not
efficiently preparable in isolation, but its size is polynomial because the device is QPT.  The advice formulation
in Definition~\ref{def:ahb-2party} permits such a state, with a negligible function that does not depend on it.  The
verifier samples $K_i$ only after this history, so the conditional state is independent of the fresh key.  The
collective-AHB reduction and the single-round operator argument therefore apply to it.  Eve's purifying system
remains side information in Theorem~\ref{thm:single-round} and is not supplied to the computational reduction.
\end{proof}

\subsection{Derivation of the occurrence factor}

Fix a one-round input with aggregate conditional test success $\omega$.  Theorem~\ref{thm:single-round} and the
definition of $g_\beta$ give
\begin{equation}\label{eq:appendix-one-round-preimage}
 g_\beta(\omega)-\xi_{\rm AHB}(\lambda)
 \le H(\widehat\Pi_i|Y_iK_i,T_i=0,R').
\end{equation}
The event $T_i=0$ consists of all generation rounds and the preimage-test rounds, so its probability is
$1-\beta\gamma$.  Removing $Y_i$ from the conditioning, averaging over $T_i$, and using the nonnegativity of the
conditional entropy of a classical register gives
\begin{align}
 (1-\beta\gamma)\bigl(g_\beta(\omega)-\xi_{\rm AHB}(\lambda)\bigr)
 &\le(1-\beta\gamma)
 H(\widehat\Pi_i|K_i,T_i=0,R')\nonumber\\
 &\le H(\widehat\Pi_i|K_iT_iR')\nonumber\\
 &\le H(\widehat\Pi_i\widehat M_i|K_iT_iR').
 \label{eq:appendix-occurrence-chain}
\end{align}
Because $G_i$ is hidden while the instrument is selected and the response is produced, given $T_i$ it is independent
of the output and of $R'$.  Thus the final conditional entropy in~\eqref{eq:appendix-occurrence-chain} is unchanged
when $G_i$ is added to the conditioning.  Replacing $g_\beta$ by the globally valid affine lower bound $h_{\omega_0}$
proves Lemma~\ref{lem:min-tradeoff}.

The factor $1-\beta\gamma$ is easily missed by counting only generation rounds.  Before the tests are opened,
preimage-test outputs are produced by the same $T=0$ instrument and also contribute entropy to the complete EAT
output.  They are removed later at the cost $L_{\rm drop}$.

\subsection{Convex acceptance and affine minimization}

For an empirical score distribution $q$, the three conditions
\begin{equation}
 q(1)\ge\gamma\omega_{\rm exp}-\delta_{\rm est},\qquad
 q(\mathsf g_{\rm bad})=0,\qquad
 q(0)+q(1)\le\gamma+\delta_{\rm samp}
\end{equation}
are affine.  Their intersection with the probability simplex is therefore convex, as the EAT requires.  Since
$f_{\min}$ in~\eqref{eq:fmin-def} is affine in $q(1)$, its infimum over this polytope is a finite-dimensional linear
program.  If $a_0\ge0$ and the polytope is nonempty, then
\begin{equation}\label{eq:appendix-explicit-t}
 t=(1-\beta\gamma)\left[
 a_0\left(\omega_{\rm exp}-\frac{\delta_{\rm est}}{\gamma}\right)
 +b_0-\xi_{\rm AHB}(\lambda)\right],
\end{equation}
provided the lower win-count face is compatible with the remaining constraints.  If a different supporting line has
negative slope, the definition~\eqref{eq:t-definition}, rather than~\eqref{eq:appendix-explicit-t}, must be used.

\subsection{From the coarse output to Alice's generation string}

Condition on the public flags $T^n,G^n$ and on $\Omega_0$.  The coarse output $O^n$ consists of Alice's generation
string $Z_A$, the coarse test outputs $O_{\rm test}$, and outputs that are trivial given $T^n$.  For each fixed value
of $G^n$ consistent with $\Omega_0$, the register $O_{\rm test}$ takes at most $3^{(\gamma+\delta_{\rm samp})n}$
values, because a fixed test type has at most three coarse outcomes and $\Omega_0$ caps the number of test rounds.
The second inequality in~\eqref{eq:classical-leakage} therefore gives
\begin{equation}
 H_{\min}^{\varepsilon_s}(Z_A|O_{\rm test}K^nT^nG^nE)
 \ge H_{\min}^{\varepsilon_s}(O^n|K^nT^nG^nE)-L_{\rm drop},
\end{equation}
which proves~\eqref{eq:Ldrop}.  The register $O_{\rm test}$ determines every announced test result $W_i$.  The
remaining public test data are the announced branch bits in preimage-test rounds and the Bobs' forwarded answers in
equation-test rounds, at most $N$ bits per test round; images, preimages, trapdoors, and Alice's equation answers
are never published.  The first inequality in~\eqref{eq:classical-leakage} charges these data
by~\eqref{eq:explicit-Ltest}, and the reconciliation and verification transcript by $L_{\rm EC}$.  These bounds are
worst-case support bounds and remain valid for arbitrary correlations and non-IID behavior.

\subsection{Abort-weighted leftover hashing}

Let $C_{\rm EC}$ contain the complete reconciliation transcript, the verification tags, and a bit $F$ recording whether
the disagreement and reconciliation checks accept, and include its full support in $L_{\rm EC}$.  The entropy bound
before privacy amplification is therefore a statement conditioned on $C_{\rm preEC}C_{\rm EC}E$, including $F$.

First suppose $\Pr[\Omega_0]\geq\varepsilon_{\rm acc}$, so that~\eqref{eq:eat-complete-output} holds.  For the
purposes of the proof, sample the two-universal hash seed and evaluate the hash on every branch of $\Omega_0$,
including branches with $F=0$; the real protocol may simply suppress that output.  With
\begin{equation}
 k:=nt-v\sqrt n-L_{\rm drop}-L_{\rm test}-L_{\rm EC},
\end{equation}
the quantum leftover hash lemma~\cite{Tomamichel2011,Renner2008} gives
\begin{equation}\label{eq:appendix-leftover}
 \DeltaTr{\rho_{K_AFC_{\rm preEC}C_{\rm EC}E|\Omega_0}}
 {\tau_{K_A}\otimes\rho_{FC_{\rm preEC}C_{\rm EC}E|\Omega_0}}
 \le 2\varepsilon_s+\frac12\,2^{-(k-\ell)/2},
\end{equation}
and the choice~\eqref{eq:key-length} makes the second term at most $\varepsilon_{\rm PA}$.  Now apply the classical
map that outputs the key only when $F=1$ and outputs an abort symbol otherwise.  Since $\Omega\subseteq\Omega_0$,
contractivity of the subnormalized trace distance turns~\eqref{eq:appendix-leftover} into
\begin{equation}\label{eq:appendix-abort-weighted}
 p_\Omega\,
 \DeltaTr{\rho_{K_ACE|\Omega}}{\tau_{K_A}\otimes\rho_{CE|\Omega}}
 \le2\varepsilon_s+\varepsilon_{\rm PA}.
\end{equation}
If instead $\Pr[\Omega_0]<\varepsilon_{\rm acc}$, then $p_\Omega\leq\Pr[\Omega_0]<\varepsilon_{\rm acc}$, and the
left-hand side of~\eqref{eq:appendix-abort-weighted} is at most $\varepsilon_{\rm acc}$ because trace distance is at
most one.  In both cases the protocol is $(\varepsilon_{\rm acc}+2\varepsilon_s+\varepsilon_{\rm PA})$-secret.  This
argument is why the final disagreement and reconciliation checks need not be inserted artificially into the
single-round score: their public records and acceptance flag are charged before the virtual extraction, and the
final filtering is an ordinary classical post-processing.

\subsection{Correctness}

Let $F_j$ be the event that Bob $j$ accepts verification although $\widetilde Z_j\ne Z_A$.  If each verification
construction fails with probability at most $\varepsilon_{{\rm ver},j}$, then
\begin{equation}
 \Pr[\Omega\wedge\exists j:\widetilde Z_j\ne Z_A]
 \le\sum_{j=2}^{N}\varepsilon_{{\rm ver},j}.
\end{equation}
Choosing the tags so that this sum is at most $\varepsilon_{\rm corr}$ proves~\eqref{eq:correctness-definition}.
Privacy amplification is deterministic given the public seed, so equal reconciled strings produce equal final keys.

\subsection{Proof of the finite-size security theorem}

\begin{proof}[Proof of Theorem~\ref{thm:main}]
Fix an arbitrary sequential strategy of the joint device.  For each round and every preceding classical history,
Lemma~\ref{lem:fresh-key-history} makes the incoming memory valid advice for the collective-AHB reduction, so
Theorem~\ref{thm:single-round} applies to the conditional response instrument.  Minimizing over the unobserved split
between the preimage and equation tests and taking the affine supporting line yields Lemma~\ref{lem:min-tradeoff}.

Lemma~\ref{lem:appendix-eat-channels} verifies the channel, reconstruction, and Markov premises of the EAT.  The
acceptance set~\eqref{eq:accepted-frequency-set} is an intersection of affine half-spaces with the probability
simplex and is therefore convex.  If $\Pr[\Omega_0]\geq\varepsilon_{\rm acc}$, applying the EAT with $f_{\min}$ and
minimizing over this set gives~\eqref{eq:eat-complete-output} for the full sequence $R_0,\ldots,R_n$; no step
replaces the carried memories by independent copies.

The chain rule~\eqref{eq:classical-leakage} first conditions on the coarse test outputs at cost $L_{\rm drop}$.  The
support bounds~\eqref{eq:explicit-Ltest} and $L_{\rm EC}$ then charge the announced test data, the reconciliation
syndrome, and the verification data, giving~\eqref{eq:main-final-entropy}.  The virtual extraction
argument~\eqref{eq:appendix-leftover}, the key-length condition~\eqref{eq:key-length}, and the final acceptance
filter give~\eqref{eq:appendix-abort-weighted}.  Together with the case $\Pr[\Omega_0]<\varepsilon_{\rm acc}$, this
proves $(\varepsilon_{\rm acc}+2\varepsilon_s+\varepsilon_{\rm PA})$-secrecy.  Finally, the correctness argument above
and~\eqref{eq:verification-correctness} give correctness error at most $\varepsilon_{\rm corr}$.  This proves all
claims of Theorem~\ref{thm:main} against arbitrary coherent attacks.
\end{proof}

\section{Honest-Noise Calculations}\label{app:noise}

This appendix derives the formulas of Section~\ref{sec:discussion} from Pauli expectations and proves the
corresponding finite-noise completeness statement.

\subsection{Depolarizing noise as classical outcome flips}

The single-qubit depolarizing channel~\eqref{eq:depolarizing-channel} acts in the Heisenberg picture as
\begin{equation}
 \mathcal D_p^\dagger(I)=I,\qquad
 \mathcal D_p^\dagger(P)=(1-p)P
\end{equation}
for $P\in\{X,Y,Z\}$.  A measurement of a Pauli eigenstate in the same basis is therefore flipped with probability
$p/2=Q$; equivalently, its correlation expectation is multiplied by $1-2Q=1-p$.

Let $e_j\in\{0,1\}$ denote the classical flip on branch $j$.  In our model $e_1=0$ and $e_2,\ldots,e_N$ are independent
Bernoulli-$Q$ variables.  The Hadamard parity remains correct exactly when $\oplus_{j=2}^Ne_j=0$.  The character
identity
\begin{equation}
 \mathbb E[(-1)^{\oplus_{j=2}^{N}e_j}]
 =\prod_{j=2}^{N}\mathbb E[(-1)^{e_j}]
 =(1-2Q)^{N-1}
\end{equation}
implies
\begin{equation}
 \Pr\!\left[\bigoplus_{j=2}^{N}e_j=0\right]
 =\frac{1+(1-2Q)^{N-1}}2.
\end{equation}
Depolarizing noise followed by a Hadamard-basis measurement is equivalent to the noiseless measurement followed
by the independent classical flips $e_j$ of the outcomes $u_j$.  The noisy test therefore passes whenever the
noiseless test passes and $\oplus_{j=2}^Ne_j=0$.  These two events are independent, and the first has probability
at least $1-\nuG$ by Proposition~\ref{prop:completeness}, so
$\omega_m^{(N)}(Q)\geq(1-\nuG)\frac{1+(1-2Q)^{N-1}}2$, which implies~\eqref{eq:omega-m-depol}.  The
preimage register and Alice's branch are untouched, which gives~\eqref{eq:omega-p-depol}.

All-party computational-basis agreement requires $e_2=\cdots=e_N=0$, and hence has probability $(1-Q)^{N-1}$.  For a
fixed Bob $j$, Alice and Bob disagree exactly when $e_j=1$, which gives quantum bit error rate $Q$.  Notice that an
even nonzero number of flips passes the equation test but destroys all-party agreement.  This is why the parity
statistic cannot itself be used as a correctness estimate.

\subsection{Low-noise expansions}

For fixed $N$ and small $Q$,
\begin{align}
 (1-Q)^{N-1}
 &=1-(N-1)Q+\binom{N-1}{2}Q^2+O(N^3Q^3),\\
 \frac{1+(1-2Q)^{N-1}}2
 &=1-(N-1)Q+(N-1)(N-2)Q^2+O(N^3Q^3).
\end{align}
The first-order coefficients coincide.  The larger second-order coefficient of the parity success records the
accepted two-error patterns.

\subsection{Finite-noise completeness}

Put
\begin{equation}\label{eq:appendix-noisy-aggregate-win}
 \underline\omega_N(Q)=(1-\beta)+\beta\left(\frac{1+(1-2Q)^{N-1}}2-\nuG(\lambda)\right),\qquad
 \Delta_{\rm win}(Q)=\gamma(\underline\omega_N(Q)-\omega_{\rm exp})+\delta_{\rm est}.
\end{equation}

\begin{proposition}[Noisy honest completeness]\label{prop:appendix-noisy-completeness}
Suppose that the transmitted links are independent and obey~\eqref{eq:depolarizing-channel}, that the honest rounds
are independent, that $\Delta_{\rm win}(Q)>0$, and that
\begin{equation}
 m_p^{\min}\leq[\gamma(1-\beta)-\delta_p]n, \qquad Q_j^{\rm tol}>Q
 \quad(j=2,\ldots,N).
\end{equation}
Then
\begin{align}
 \Pr[\neg\Omega]
 &\leq e^{-2n\Delta_{\rm win}(Q)^2}+e^{-2n\delta_{\rm samp}^2}
 +e^{-2n\delta_p^2}\nonumber\\
 &\quad+\sum_{j=2}^N e^{-2m_p^{\min}(Q_j^{\rm tol}-Q)^2}
 +\varepsilon_{\rm post},\label{eq:appendix-noisy-completeness-bound}
\end{align}
where $\varepsilon_{\rm post}$ bounds the failure probability of honest reconciliation, verification, and privacy
amplification.
\end{proposition}

\begin{proof}
Let $X_i$ indicate a passing test, $Y_i$ a test round, and $Z_i$ a preimage-test round.  Their means are at least
$\gamma\underline\omega_N(Q)$, exactly $\gamma$, and exactly $\gamma(1-\beta)$.  Hoeffding's inequality applied at the
three protocol thresholds gives the first three terms in~\eqref{eq:appendix-noisy-completeness-bound}.

Conditioned on the set of $m_p$ preimage-test rounds, Alice and Bob $j$ disagree independently with probability $Q$.
On $m_p\geq m_p^{\min}$, Hoeffding's inequality bounds the probability that their empirical disagreement exceeds
$Q_j^{\rm tol}$ by $e^{-2m_p^{\min}(Q_j^{\rm tol}-Q)^2}$.  A union bound over the Bobs and the post-processing failure
event proves the proposition.
\end{proof}

\subsection{Why the same convexified curve must be used}

At the honest point, the two test success probabilities are
\begin{equation}
 a=1,\qquad
 b\geq\frac{1+(1-2Q)^{N-1}}2-\nuG(\lambda).
\end{equation}
The protocol observes only $(1-\beta)a+\beta b$.  A malicious device with the same aggregate value may realize a
different pair $(a',b')$.  The definition~\eqref{eq:g-beta} takes the infimum over all such pairs before
convexification.  Consequently, the sound asymptotic substitution is
\begin{equation}
 \underline g_\beta((1-\beta)a+\beta b),
\end{equation}
not $g(a,b)$, and not the uncertainty expression obtained by setting $c=1/2$.  This remains true even when the
honest device itself is IID: device-independent soundness must cover every adversarial realization compatible with
the accepted statistic.

\section{Operational Form of Everlasting Security}\label{app:lifting}

We finish by restating security lifting as a distinguishing experiment.  This makes clear both the strength of the
claim and why it does not follow from an ordinary computational group key exchange definition.

Let a bit $B$ choose between the real accepting state $\rho_{K_ACE|\Omega}$ and the ideal state
$\tau_{K_A}\otimes\rho_{CE|\Omega}$.  For any measurement performed after an arbitrary channel $\Gamma$, the optimal
distinguishing advantage is the trace distance between the two resulting states.  Proposition~\ref{prop:lifting}
therefore implies the same abort-weighted advantage bound $\varepsilon_{\rm sec}$ for every such experiment, including
unbounded measurements and channels.

The weighting by the acceptance probability can be seen directly.  Define the subnormalized accepting states
\begin{equation}
 \rho^{\rm acc}_{K_ACE}:=p_\Omega\rho_{K_ACE|\Omega},\qquad
 \sigma^{\rm acc}_{K_ACE}:=p_\Omega\tau_{K_A}\otimes\rho_{CE|\Omega}.
\end{equation}
Then the secrecy premise is
$\frac12\|\rho^{\rm acc}_{K_ACE}-\sigma^{\rm acc}_{K_ACE}\|_1\le\varepsilon_{\rm sec}$.  For any completely positive
trace-nonincreasing map $\mathcal F$ describing a later event, data processing gives
\begin{equation}\label{eq:subnormalized-lifting}
 \frac12\left\|\mathcal F(\rho^{\rm acc}_{K_ACE})
 -\mathcal F(\sigma^{\rm acc}_{K_ACE})\right\|_1
 \le\varepsilon_{\rm sec}.
\end{equation}
Thus an adversary cannot amplify a rare accepting or postselected branch by discarding its probability.  If one
renormalizes a later event of probability $p_F$, the corresponding conditional distance can be as large as
$\varepsilon_{\rm sec}/p_F$; the operational statement is the weighted bound~\eqref{eq:subnormalized-lifting}.  This
is the same reason the factor $p_\Omega$ appears in the base security definition.

Suppose that years after the execution an algorithm $\mathcal L$ solves the LWE instances represented in the public
transcript.  The composed map
\begin{equation}
 \Gamma_{\mathcal L,\mathcal U}
 :=\mathcal U\circ(\mathcal L\otimes\operatorname{id}_{K_AE})
\end{equation}
may first run this solver and then use its output, the key, and Eve's quantum memory in any application
$\mathcal U$.  Contractivity applies to this composition exactly as to every other channel; no
efficient-distinguisher qualifier reappears after~\eqref{eq:lift-premise}.

By contrast, a conventional computational group key exchange premise has the form
\begin{equation}
 \left|\Pr[\mathcal A(\mathsf{Real})=1]
 -\Pr[\mathcal A(\mathsf{Ideal})=1]\right|\le\negl(\lambda)
\end{equation}
only for efficient $\mathcal A$.  Contractivity of trace distance cannot be invoked, because the premise is not a
trace-distance bound.  Once $\mathcal L$ makes the relevant distinguishing or key-recovery task efficient, it falls
outside the original hardness assumption and the guarantee does not lift.  This logical difference, rather than a
claim about the relative plausibility of lattice assumptions, is the basis of the comparison in
Section~\ref{sec:pq-gke-comparison}.

Finally, the ideal-world comparison follows the same higher-level key-use map as the real world.  If an application
intentionally reveals $K_A$, the ideal application reveals an independent uniform key as well; the contractivity
statement remains true but no longer expresses secrecy from the recipient of that intentional release.  Similarly,
adding previously excluded device memory or unerased verifier state changes the input systems and is not justified
by Proposition~\ref{prop:lifting}.

The order of quantifiers is therefore
\begin{gather}
 \bigl[\text{LWE hard while the QPT device executes}\bigr]
 \Longrightarrow
 \bigl[\rho^{\rm acc}_{K_ACE}\approx_{\varepsilon_{\rm sec}}
       \sigma^{\rm acc}_{K_ACE}\bigr]\nonumber\\
 \Longrightarrow
 \bigl[\text{security against every later channel}\bigr].
\end{gather}
The first implication is a theorem under a computational assumption; the second is unconditional data processing.
Breaking LWE later cannot alter the density operator that has already been produced.  What it can do is enable a new
channel on the registers that survived the execution, and that possibility is already universally quantified in the
final implication.

\end{document}